\documentclass[12pt,preprint]{aastex}

\usepackage{emulateapj5}
\usepackage{epsf}
\usepackage{graphics}

\newcommand{\kms}{km\,s$^{-1}$}
\newenvironment{inlinefigure}{
\def\@captype{figure}
\noindent\begin{minipage}{0.999\linewidth}\begin{center}}
{\end{center}\end{minipage}\smallskip}

\slugcomment{Accepted to the Astronomical Journal}

\shorttitle{Defining Cluster Populations}
\shortauthors{Conselice, Gallagher \& Wyse}

\def\deg{$^{\circ}\,$}
\def\solm{M$_{\odot}\,$}
\def\kms{km s$^{-1}$}

\begin{document}

\title{Galaxy Populations and Evolution in Clusters II: Defining
Cluster Populations}

\author{Christopher J. Conselice$^{1,2,3}$,  John S. Gallagher, III$^{3}$, 
Rosemary F.G. Wyse$^{4,5}$}

\altaffiltext{1}{California Institute of Technology, Mail Code 105-24, Pasadena, CA}

\altaffiltext{2}{Space Telescope Science Institute, 3700 San Martin Drive, Baltimore, MD}
 
\altaffiltext{3}{Department of Astronomy, University of Wisconsin, Madison 
475 N. Charter St. Madison, WI}

\altaffiltext{4}{School of Physics \& Astronomy, University of St.~Andrews,
Scotland}

\altaffiltext{4}{Permanent address: Department of Physics \& Astronomy, The Johns Hopkins University}

\begin{abstract}

This paper presents quantitative techniques for studying, in an
unbiased manner, the photometric and structural properties of galaxies
in clusters, including a means to identify likely background objects
in the absence of redshift information.  We develop self-consistent
and reproducible measurements of fundamental properties of galaxies
such as radius, surface brightness, concentration of light and structural
asymmetry.  We illustrate our techniques through an application to deep UBR
images, taken with the WIYN 3.5m telescope, of the central
$\sim$~173~arcmin$^{2}$ (or 0.3 Mpc $\times$ 0.3 Mpc) of the cluster Abell 
0146 (Perseus).  Our
techniques allow us to study the properties of the galaxy population
in the center of Perseus down to M$_{B} = -11$.  Using these 
methods, we describe and characterize a well-defined
relation between absolute magnitude and surface brightness for galaxy
cluster members across the entire wide range of galaxy luminosity from
M$_{B} = -20$ to M$_{B} = -11$ independent of galaxy type.
The galaxies that are assigned by our
techniques to the background show no such tight relationship
between apparent magnitude and surface brightness, with the exception
of those we identify as being members of a background cluster of galaxies at 
$z \sim 0.55$. 
We, however, find that at the fainter magnitudes, M$_B > -16$, there is
a large scatter about the underlying color--magnitude relation defined
by the brighter galaxies.   Our analysis of galaxies at the center of the 
Perseus cluster further indicates
that the vast majority are `normal', with little evidence for structural 
or photometric properties associated with active evolution; we however 
discuss the detailed properties of a handful of unusual galaxies.  
Finally, the galaxy luminosity function of the Perseus
cluster center is computed, with a derived faint end slope of $\alpha =
-1.44\pm0.04$, a value similar to those previously obtained for other
nearby rich galaxy clusters.

\end{abstract}


\section{Introduction}

The oldest (mean ages $>$ 10~Gyrs) and most apparently quiescent
galaxies, without obvious evidence for recent star-formation or interactions
with other galaxies, in the local Universe are giant ellipticals found in rich
clusters of galaxies. 
These rich clusters contain a wide range of
galaxy types, including a large population of low-mass systems
(Ferguson \& Binggeli 1994), whose history may well be very different
from that of the giant ellipticals (Conselice, Gallagher \& Wyse 2001a;
hereafter Paper I, and references therein).   Investigations across the range
of galaxy populations in an individual cluster can provide insight into the
past
evolution of galaxies in these extreme environments, and can constrain 
on-going processes. 
Both bright  and faint galaxies in the nearby Coma and
Virgo clusters are now well studied and characterized (e.g., Binggeli,
Sandage \& Tammann 1985; Thompson \& Gregory 1993; Secker, Harris \&
Plummer 1997; Vazdekis et al.~2001; Paper I) 
revealing basic photometric and spectroscopic properties of
these systems. However, the nearby Bautz-Morgan class II-III cluster 
Perseus (Abell
462) with a redshift V$_{r}$ = 5366 \kms (Struble \& Rood 1999) and
distance\footnote{Using H$_{0}$ = 70 km s$^{-1}$ Mpc$^{-1}$}, D = 77
Mpc, has not yet been studied in detail, partly due to its low Galactic
latitude ($b \sim -13^o$).  This paper is a first step towards such a
study.

In addition to being one of the nearest rich clusters, Perseus
is exceptional in several ways.  It is the brightest observed
X-ray cluster (e.g. Nulsen \& Fabian et al.~1980; Ulmer et al.~1980)
with possibly a large cooling flow centered on the extraordinary
galaxy NGC~1275 (e.g. Allen \& Fabian 1997).  The Perseus cluster also
has one of
the highest known internal cluster velocity dispersions, $\sigma = 1260$ \kms
(Kent \& Sargent 1983), and a strong morphological segregation, with few
spiral galaxies found in its densest regions (e.g., Brunzendorf \&
Meusinger 1999).  The peculiar central galaxy, NGC~1275 (Perseus A),
is a strong radio source, has a non-thermal active nucleus and unusual
stellar components, as well as a spectacular system of optical
emission line filaments (e.g., Conselice, Gallagher \& Wyse 2001b, and
references therein).  As argued in Conselice et al.~(2001b), a recent
merger/accretion of a small group of galaxies into the cluster may be
responsible for several of these manifestations. 

The Perseus cluster, as revealed through
galaxy counts, has a compact core, with a size 0.1 Mpc, and 
a unimodal structure, as opposed to the double structure
found in, for example, the Coma cluster (e.g.,
Colless \& Dunn 1996).   There is  no significant spatial
clumping in the distribution of brighter cluster galaxies, although the 
nearly linear array of galaxies near its center may signify the presence of 
weak substructure.
The galaxies in the Perseus cluster also have a smooth gaussian-like 
velocity distribution (Giradi et al.~1997), although both
Slezak, Durret, \& Gerbal (1994) and Mohr, Fabricant, \& Geller (1993)
find some evidence of substructure in the X-ray emission.  
Perseus also harbors an
exceptionally large population of radio head-tail sources, including
NGC~1275 (Sijbring \& DeBruyn 1998). These sources are possibly the
result of interactions between the host galaxies and the intracluster
medium, with the large cluster velocity dispersion resulting in stronger ram
pressure forces ($\propto v^2$) compared to a more typical cluster with slower
relative velocities between galaxies and the ICM.   These extreme conditions
make the core region of Perseus unique in the nearby universe and could
have a strong effect on the state and evolution of its galaxies.  However,
the same basic physical mechanisms should be occurring in in all clusters and
they are perhaps faster or stronger in Perseus, making evolutionary  
effects easier to study.

However, as the Perseus cluster is one of the richest systems in the nearby
universe, and is at a distance significantly less than that of the Coma
cluster (Baum et al.~1997), it represents a good opportunity to study
the signatures of dynamical and photometric evolution of galaxies in
such environments, including the intrinsically faint members.  The
development of the necessary tools for such a program is the aim of
the present paper, while a companion paper (Conselice et al. 2002; hereafter
Paper III) presents details of the results of an investigation of faint
Perseus cluster members.

In this paper we define global photometric and structural parameters for
galaxies, taking care to adopt measuring  techniques which will be useful 
for comparisons between galaxy clusters.   For example, 
commonly used isophotal magnitudes are ill-suited for
comparing galaxies observed in different conditions, and especially at
different distances where the ratio of metric to isophotal radii can
vary strongly.  As a result, the use of
isophotal radii can bias studies of both galaxy evolution and
cosmology (Petrosian 1976; Sandage \& Perelmuter 1990; Dalcanton
1998).  We therefore adopted Petrosian
radii to measure sizes of galaxies and to define photometric apertures.

Additionally, it has
become clear recently that structural parameters can reveal much about
the evolutionary state of galaxies (e.g., Abraham et al.~1994; Conselice 1997;
Conselice, Bershady \& Jangren 2000; Bershady, Jangren \& Conselice 2000).
Attempts have also been made for some time  to classify individual galaxies in 
clusters
using structural parameters (e.g., Ichikawa, Wakamatsu \& Okamura
1986), although no systematic classification system has yet been presented.
This paper presents the framework of such a system in clusters (see \S 4.3).  
We developed techniques to distinguish between cluster and background 
galaxies, as well as methods for quantifying galaxy structures using
the asymmetry and concentration indexes.  
In galaxies there are essentially two `ages' related to the star-formation
history and to the mass assembly history respectively, and photometry helps
decipher the former, and structure the latter.

Photometry of bright cluster galaxies has revealed the existence of a
well-defined correlation between the color and magnitude  of
elliptical galaxies (e.g., Kormendy 1977; Bower, Lucey \& Ellis 1992), such 
that fainter galaxies are bluer.  This is normally interpreted as a
metallicity effect, with redder ellipticals being more metal-rich
(e.g. Larson 1974; Kauffman \& Charlot 1998; Vazdekis et al. 2001).  
The data we present
here allow us to investigate such scaling relations, and others
between the photometric and structural parameters down to very faint
intrinsic luminosities, M$_{B} \sim -11$. 
These faint low-mass galaxies appear to be dwarf
ellipticals; we however denote these galaxies using the more general term
`low-mass cluster galaxy (LMCG)' (a nomenclature we  adopt also in Paper 
III), so as to not bias
the interpretations of what these objects are.
We also compare observed UBR color-color diagrams with the predictions for
old stellar populations with a range of
metallicities, and find results that support the interpretation that 
the range of colors of ellipticals in Perseus can
be explained by metallicity differences (see Paper III for a full analysis).

This paper is organized as follows: in \S 2 we describe the data and
analysis techniques, \S 3 is a discussion of background galaxy
contamination, \S 4 describes an analysis of the data set, and \S 5
contains a summary.  We assumed a  distance to the Perseus cluster 
of 77 Mpc throughout this paper, giving a scale of $\sim$20~kpc per arcmin.

\section{Observations and Reductions}

  All of the imaging data used in this paper, and in Paper III, were taken
with the WIYN 3.5m f/6.2 telescope located on Kitt Peak.   The B and R images
were acquired with a thinned
2048$^{2}$ pixel S2kB charged coupled device (CCD).  The scale for these
images is 0.2\arcsec\, pixel$^{-1}$, with a field of view 6.8\arcmin $\times$
6.8\arcmin.   The imaging
took place on the nights of 1998 November 14 and 15, under photometric
conditions.  The average seeing for all images was 0.7\arcsec. 
The images of the Perseus cluster were taken in four different fields
towards the center of the cluster.  Harris B and R, as well as narrow band 
H$\alpha$, filters were used on each of the four fields.  The exposure times
were 2400~sec in R, and 2000~sec in B. The H$\alpha$ images were 
discussed in  Conselice et al.~(2001b).  Figure~1
shows a mosaic image of the area covered in this paper and in Paper~III.
 
  Ten flat fields were obtained in each filter prior to each night of
observing with the S2kB.  A single flat field for each filter for each
night was created by taking the median of its respective flats.  A
constant zero level pedestal was determined from the over-scan region,
and subtracted, rather than using standard bias frames, since it was
discovered that adopting the latter technique increases the
uncertainty in photometry obtained from this CCD.

To perform accurate photometry on the galaxies in these fields, we
align and transform the B images to match the R images.  A simple
linear interpolation was found to be inadequate, as the point spread function
(PSF) for each filter varied
differently with position on the CCD, and small scale differences also
exist. We therefore used a non-linear 2nd
order fitting routine to
match the positions of objects, including corrections for slight
rotations between the B and R frames.  As a result, the B and R frames
are matched to within an rms of 0.2 pixels, or 0.04\arcsec.

We also obtained UBR-band images with the WIYN Mini-Mosaic camera for
two 9.6\arcmin\, $\times$ 9.6\arcmin\, fields around the center of the
four combined S2kB fields.  Mini-Mosaic consists of two SITe 4096
$\times$ 2048 CCDs separated by a gap of 5\arcsec\, and with a scale of
0.14\arcsec\, pixel$^{-1}$. Exposure times for the U band images are
2400 sec per field, resulting in good photometry for the cluster
galaxies, with shorter exposure Harris B and R images taken for
calibration purposes only. The images were reduced using the IRAF
package MSC, with flat fielding and alignment done in the same manner
as for the S2kB images, but with the bias subtraction performed in the
standard manner using bias frames.  During the Mini-Mosaic imaging run
the seeing in the U band images was 0.9\arcsec. \vspace{0.2cm}

\begin{inlinefigure}
\begin{center}
\resizebox{\textwidth}{!}{\includegraphics{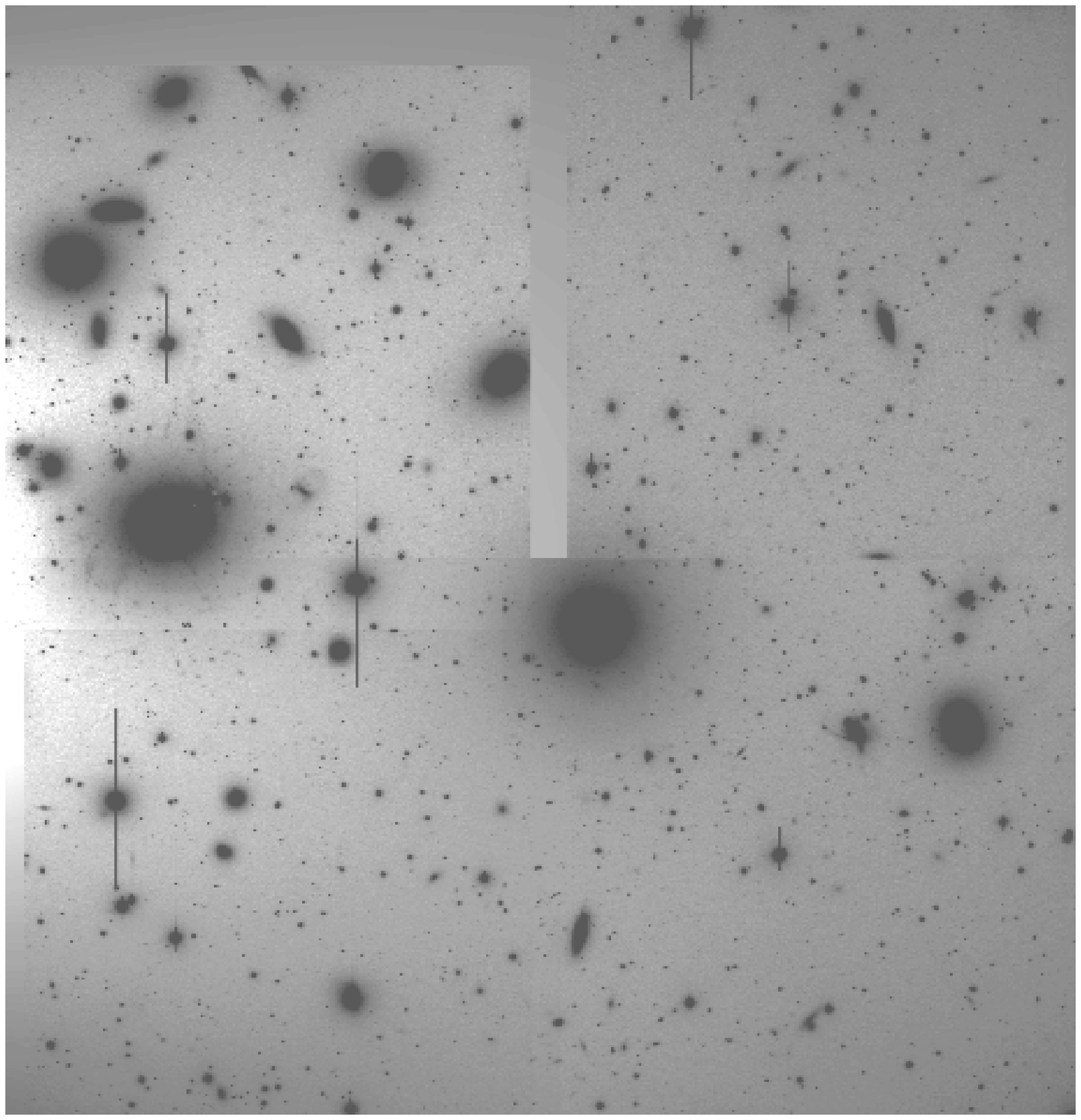}}
\end{center}
\figcaption{Mosaic image of the central 173 arcmin$^{2}$ of the Perseus
cluster imaged by WIYN in the R band and studied in this paper.  
NGC 1275 is the large galaxy in the left (east) part of the image.}
\end{inlinefigure}

\subsection{Photometry}

Images of Landolt standard star fields were taken 
throughout each night, and as 
a result, we are able to  fit accurate zero points, airmasses, and color
terms, for both nights.  We 
fit  the following photometric solution for each of the B-band  and R-band 
 S2kB data: 

\begin{equation}
F_{B,R} = f_{B,R} + a_{1} + a_{2} \times X_{B,R} + a_{3} \times (B-R),
\end{equation}

\noindent where $F_{B,R}$ is the magnitude of an object, $f_{B,R}$ is
the instrumental magnitude, $X$ is the airmass of the observation,
while $a_{1}, a_{2}, a_{3}$ are the zero point offset, airmass and
color terms.  Higher order color terms were also allowed initially,
but were found to be smaller than the photometric random errors, and
so were left out of the final photometric calibration.  The
coefficients $a_i$ were set to be constant throughout each night.  The
standard star calibration provides magnitudes with RMS random errors
of $\Delta$mag~$\sim$~0.04 in B and $\Delta$mag~$\sim$~0.03 in the R
band.  To correct for Galactic extinction, we first investigated the
amplitude of any variation over the area studied here through
examination of the COBE/DIRBE and IRAS/ISSA maps from Schlegel,
Finkbeiner \& Davis (1998).  These maps show that Galactic absorption
in the B-band varies by only 0.03~mag over 15\arcmin.  Most of this
variation is due to one small area of lower than average extinction.
These variations are equal to, or lower than, the random photometry
errors and are hence ignorable. For the correction, we adopted the value of 
E(B--V) = 0.171 and
$R_{V} = 3.1$ and the empirical relations of Cardelli, Clayton and
Mathis (1989).  We also correct for the slight k-correction, using the
prescriptions of Poggianti (1997).

The Mini-Mosaic data, taken 13 months later, were calibrated in the
same manner.  There is considerable overlap between the Mini-Mosaic and S2kB
fields and thus we
can make a direct comparison to determine the accuracy of 
the photometric calibration as a function of
magnitude.  Figure~2 shows the result of such a   
comparison between the B magnitudes
and (B$-$R) colors for the galaxies in the overlapping regions of the two
fields-of-view --  the photometry,
carried out on two completely different detectors, over a year apart,
and with different standard star calibrations, is stable and robust.  
The scatter of the difference is low for the brightest
galaxies, RMS $\sim$ 0.05 mag, while the fainter galaxies have a
larger scatter.  Often the most deviant points are due to residual stellar
light contamination in the aperture used (see 
\S 2.2 below for a discussion of our techniques to remove stars). 

\subsection{Galaxy Detection and Star Removal}

We detected both galaxies and stars with the FOCAS software package
(Jarvis \& Tyson 1981).  A detection threshold of 5$\sigma$ above the
sky is used to find only modestly faint galaxies and not the vast
number of faint, low surface brightness galaxies, which are probably mostly
background objects.  We furthermore used a strict deblending criterion
to remove detections of globular clusters often seen near many of
the giant galaxies.  The deblending parameters adopted are appropriate for 
the case at hand, in which most of the objects are early-type galaxies
without significant substructure, and the routine does not  
split single galaxies into several spurious separate detections.  FOCAS  does 
not detect the
brightest galaxies, thus these are added in by hand, as are any
obvious galaxies that FOCAS misclassified as stars due to their bright
nuclei. After removing objects classified by FOCAS as stellar, 
and adding in the brighter
galaxies, our final catalog contains 904 galaxies over the 173 arcmin$^{2}$
of the survey.  As we argue below, most of these objects are background
galaxies.  

Photometry of galaxies in the Perseus cluster is complicated
by its low Galactic latitude, and the resulting presence of
numerous foreground Galactic stars. We subtracted the occasional overlapping
stellar images prior to obtaining photometry for the galaxies using the
point-spread function fitting and subtraction routines in the IRAF DAOPHOT
package. The stellar PSF was characterized as a function of magnitude
empirically, from measurements of stellar images of different 
magnitudes spread across the field.   It proved possible 
to remove all the light from most stars to within 1\%.  We then used
these star-subtracted images for all further analysis of the  
photometry and structural parameters of Perseus galaxies. 

\begin{inlinefigure}
\begin{center}
\resizebox{\textwidth}{!}{\includegraphics{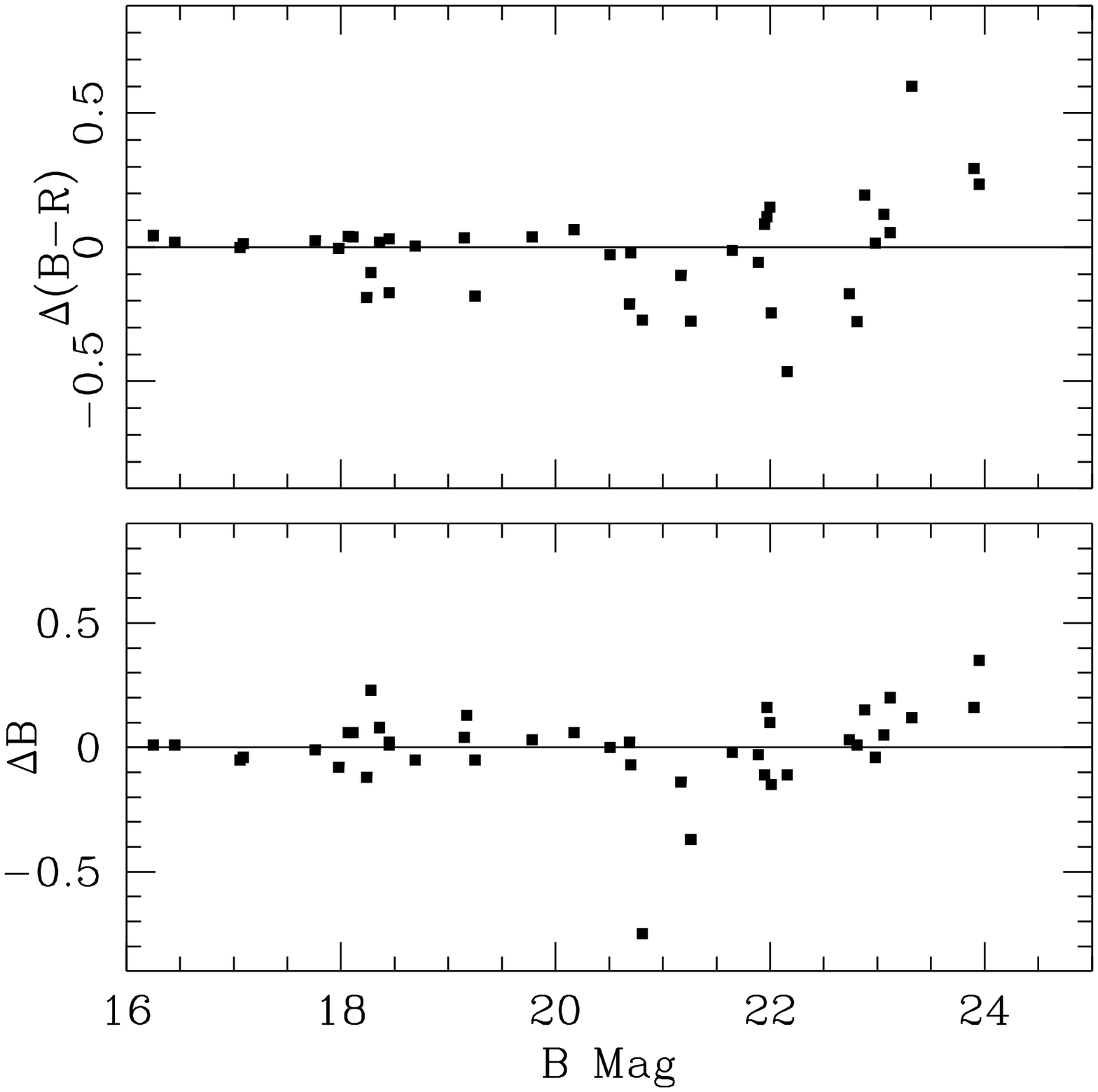}}
\end{center}
\figcaption{Difference in measured (B$-$R) color and apparent B 
magnitudes for galaxies observed with both the Mini-Mosaic and S2kB 
CCD cameras on the WIYN telescope.  The Mini-Mosaic data are not as
deep in the B and R bands as they are in the S2kB images.}
\end{inlinefigure}

\subsection{Different Measurements of Radius}

Defining a galaxy's radius is not trivial, and many methods exist,
including the traditional and popular approach of measuring a radius out to an
isophotal level, such as the $\mu_{pg}$ = 26.5 mag arcsec$^{-2}$
Holmberg radius.  While almost any consistently measured `radius' would
be adequate for this study, it would be inaccurate, and perhaps
irresponsible, to use radii that are not reproducible in other
clusters and galaxies at various distances, something which isophotal
radii are ill-suited to do.  Even so-called metric radii that are
based on the percentage of light within an aperture are often based on
``total'' light measurements that are isophotally biased.

To overcome these difficulties, we use the inverted form of the
Petrosian radius (Petrosian 1976; Kron 1995).  The inverted Petrosian
parameter, $\eta$, is the ratio of the local surface brightness
$l(\theta)$ at a given angular distance, $\theta$, from the center of
the galaxy to the mean surface brightness within $\theta$, assuming circular 
apertures and azimuthal averaging:

\begin{equation}
\eta = \frac{1}{2} \frac{{\rm d ln} l(\theta)}{{\rm d ln} \theta} = \frac{I_{\theta}}{<I>_{\theta}}.
\end{equation}

\noindent The inverted Petrosian radius parameter is defined as the projected 
radius at
which $\eta$ equals some value between 1 and 0.  In general, the lower the
value of $\eta$ the larger the value of the Petrosian radius.   
The Petrosian radii were found by
Bershady et al.~(2000) and Conselice et al.~(2000a) to be the most
consistent and robust measures of radius to use in the derivation of
structural parameters of galaxies and have been used for cosmological
and galaxy evolution work (e.g., Sandage \& Perelmuter 1990; 
Sandage \& Lubin 2001;  Blanton et al. 2001).

We used the R$_{p}$ = 3~$\times$~r($\eta$ = 0.5) radius to measure the
total light in a galaxy from which half-light radii, `total'
magnitudes and colors are derived.  The R$_{p}$ radius defined in this
way has values similar to the $\eta=0.2$ radius where $\sim$99\% of the
light of most galaxies is contained (Bershady et al.~2000).  We chose the 
present definition since for faint
galaxies, which are often barely resolved, as in our case, the derivation
of the $\eta = 0.2$ radius often never converges, and experimentation
revealed that $\eta = 0.5$ was the most reliable physical radius that
corresponded well to subjective eye estimates of galaxy sizes.  Further
experimentation with a high redshift cluster, MS-1054, confirms the
reliability of R$_{p}$ for measuring metric radii (Conselice et
al. 2002 in preparation).

\subsection{Photometric Errors and Completeness}

This paper includes photometry of faint galaxies which have magnitudes
m$_{B} <$ 25 within R$_{p}$.   Each galaxy has an associated random error
based on its individual photometric measurement within R$_{p}$. We performed
limited Monte-Carlo simulations to determine how accurate
these estimated random errors are, and to search for any systematic
errors that may be present across our images.

To do this we created 20 simulated galaxies, multiple times,  all at the same
magnitude and
place them randomly in the Perseus fields.  These simulated galaxies range in
magnitude in B and R from 18.5 to 24.5, all with exponential profiles that
match dwarf galaxies, and 
half-light radius (equal to 1.7 exponential scale-lengths) similar to a
typical Perseus low mass cluster galaxy with a half-light radius
of 2\arcsec.  
This size also ensures that the surface brightness
of the simulated objects is similar to that of the faintest objects in our 
study. Each galaxy in the simulations has its magnitude measured following the
same techniques and methods used to compute magnitudes
of actual Perseus galaxies, as outlined above.

Figure~3 shows the results of these simulations for retrieved magnitudes
and colors.  The triangles in Figure~3a
show the average difference between measured and input magnitudes as a
function of input magnitude in the B-band image,
while boxes represent the average differences in the R-band image.  The
error bars in Figure~3 represent the 1$\sigma$ variation of the
differences.  Figure~3b shows the average retrieved color differences and
their 1$\sigma$ variations.  These simulations show that it is possible to
measure very reliably the
magnitudes and colors of objects down to an absolute magnitude M$_{B}$
$\sim -12.5$ at the assumed distance and with reddening accounted for.  At 
M $\sim -12.5$ the mean uncertainties are:  $<({\rm R_{sim}} - 
{\rm R_{meas}})>$ = 
0.04$\pm$0.03, $<({\rm B_{sim}} - {\rm B_{meas})}>$ = 0.03$\pm$0.01 and
$<{\rm ((B-R)_{sim}} - {\rm (B-R)_{meas}})> = -0.02 \pm0.03$. 
These are
similar to or less than the observational errors at our magnitude limit.
There are therefore no strong systematic errors in measurements down to
${\rm M} = -12.5$, and the reported photometric random errors are likely 
accurate.  We can reach somewhat fainter limits but
with an increased systematic uncertainty.

\begin{inlinefigure}
\begin{center}
\vspace{-2cm}
\hspace{-2.5cm}
\rotatebox{-90}{
\resizebox{\textwidth}{!}{\includegraphics[bb = 25 25 625 625]{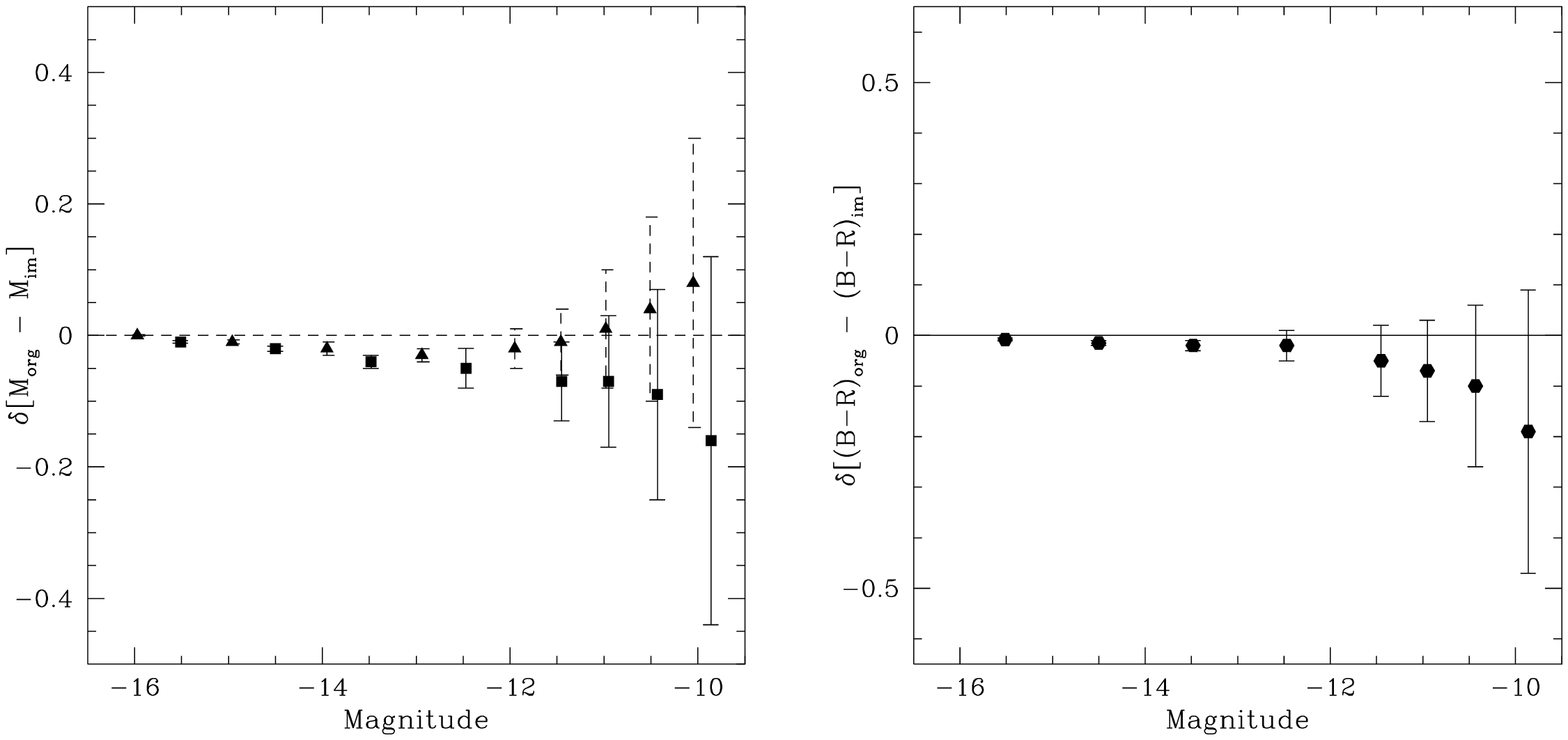}}}
\end{center}
\vspace{-2cm}
\figcaption{Results of Monte Carlo simulations performed by placing 
simulated galaxies in the images used in this paper.  The left
panel shows the resulting magnitude difference between the original and input
magnitude, M$_{\rm im}$ as a function of the input magnitude, converted
into absolute values assuming these objects are at the distance
of the Perseus cluster.  The squares represent
the simulations done in the R band, while the triangles are for the B band.
The error bars are formal 1$\sigma$ variations of the average differences.  
The right panel shows the difference between the input and output
colors.  These results support the validity of our data reduction
procedures for objects down to M$_{B} = -11$.}
\end{inlinefigure}

Figure~4 is a  plot of the number counts for all galaxies detected
by FOCAS in the sample area (173 arcmin$^{2}$).  The number counts
turn over fainter than  m$_{B}$ = 23.5, and as such they are not complete to
magnitudes fainter than this. This corresponds to M$_{B} = -10.7$, about a
magnitude fainter than the limit reached by previous studies of the Perseus
cluster luminosity function (De Propris \& Pritchet 1998).
It is possible that we are not complete
at brighter magnitudes if the true counts are steeper than we observe.
All results pertaining to complete samples, such
as luminosity functions, will be examined at magnitudes brighter
than this limit, i.e., M$_{B} < -11$.

\subsection{Structural Parameter Definitions}

We performed photometry on each of the 904 galaxies in our sample using
the APHOT package in IRAF.  APHOT is an aperture photometry program
that allows users to chose various photometric apertures, as well as
centering and sky fitting routines.  For each galaxy, we fit the sky
using an annulus of width 10~pixels, and radius 100~pixels (20\arcsec), 
centered on the galaxy.  While we tried smaller annuli and
generally produced the same results, we favored the larger annulus
radius as our images were generally flat and the larger number of
pixels allows a better estimate of the true background and reduced the
significance of any contamination from other galaxies as more sky area
is covered.   We adopted the centroid of each galaxy's light
distribution as the center for the photometry routines.  
Total magnitudes and colors within the  R$_{p}$ radius were measured from 
the UBR
frames.  We also measured the central surface-brightness $\mu_{B}$,  
defined as the mean surface brightness within the central 2$\arcsec$ of
each galaxy.  A central color is also measured within the same radius.

We also computed for each galaxy the half-light or
effective radius, R$_{e}$, defined as the point on the curve of
growth where half of a galaxy's light (based on the total magnitude out to the Petrosian radius we adopt) is contained. 
Another photometric parameter we  compute is the concentration
index, C, defined as the ratio of the radii containing 80\% (${\rm r}_{80}$) 
and 20\% (${\rm r}_{20}$) of a
galaxy's light (e.g., Kent 1985; Bershady et al.~2000),

\begin{equation}
{\rm C} = 5 \times {\rm log} \left(\frac{{\rm r}_{80}}{{\rm r}_{20}}\right).
\end{equation}

\noindent A higher value of C implies that the light in that galaxy is more
concentrated towards the center, and likewise, a lower C value indicates the
light is less concentrated.  Typically early-type galaxies have
the highest C value, while later types have lower values (Bershady
et al.~2000).   Within the R$_{p}$ radius a simulated galaxy with an r$^{1/4}$
surface brightness profile has a concentration value C = 3.2, while
a pure exponential disk has C = 2.5, and a constant surface brightness system has C = 1.5.

The other quantitative morphological parameter we use is the asymmetry
index $A$ (Conselice 1997, Conselice et al.~2000a), a parameter that
measures the deviation of a galaxy's light from perfect 180\deg\, symmetry.
Mathematically, it is defined as

\begin{equation}
A = {\rm min} [\frac{\Sigma |(I_o - I_{180})|}{\Sigma |I_{o}|}] - {\rm min}[\frac{\Sigma |(B_o - B_{180})|}{\Sigma |I_{o}|}],
\end{equation}

\noindent where $I$
represents the intensity values of the image pixels, the subscript indicating 
the rotation angle, and $B$ represents a blank region containing
background light. The second term is used to correct for
sky and noise effects.   The more disturbed the structure of a galaxy is, the
higher the measured asymmetry, $A$ (\S 4.3.1).

\subsection{Morphological Classifications}

One of the key tools for understanding galaxies is to classify 
them into distinct populations (e.g., van den Bergh 1998).  This is
especially true for the members of galaxy clusters (see Paper~I for a
longer discussion).  Although ideally we would classify galaxies
without the use of subjective eyeball classifications, it is still
necessary to do this at some level to calibrate the more objective,
automatic classification techniques and to place galaxies in rough
population classes.

We  classified galaxies in these images by carefully examining each
object, and placing them into 6 rough morphological
types.  These are: giant ellipticals, low-mass cluster galaxies (LMCGs),
early-type
spirals, late-type spirals, peculiar galaxies and background galaxies.
Lenticular (S0) galaxies were placed into the early-type spiral galaxy bin.

Giant ellipticals are defined in this paper as large, high surface
brightness, mostly symmetric,  objects without the presence of an outer disk.
Ellipticals also have high concentration values and low asymmetries (Conselice
et al.~2000a).   Early and late-type
spirals are objects with disks and/or spiral patterns.  Early-type disks
are galaxies whose bulges are roughly brighter than their disks, or with
an estimated
B/D $> 1$, while late-type disks are those where the disk light dominates the
bulge, or an estimated B/D $< 1$.
Peculiar galaxies are ones that
do not fit into the classical Hubble sequence.  In our Perseus cluster 
sample, this category includes one possible
merger, and one interesting galaxy near NGC~1275 that is symmetric, but
with an unusual structure (see \S 4.5 and Conselice, Gallagher \& Wyse 2001b),
as well as NGC~1275 itself.

We defined early-type LMCGs (see \S 1 and Paper III) as objects with low 
surface brightnesses, low light
concentrations, and exponential surface brightness
profiles.  We furthermore used the constraint that
these early-type LMCGs are symmetric
objects with no obvious signs of distortions or sub-structure.  Our
good resolution (0.7\arcsec) images allow us to make these determinations
for even small, faint objects.   Note that we only consider objects
with sizes $> 1\arcsec$ from which we can make these morphological estimates.
The summary of these criteria for early-type LMCG classification is the 
following:

\noindent (i). Total (B$-$R) color values  $<$ 2.  Galaxies redder than
this are almost always in the background, although Mobasher et al.
(2001) find confirmed faint members of the Coma cluster with colors
(B$-$R) $> 2$.

\noindent (ii). Symmetric, round or elliptical shapes, without
evidence for internal structures that might be due to 
star-formation, spiral structures, or other internal features.  Background 
galaxies are often morphologically disturbed and
can be identified in high resolution images (e.g., Conselice 2001).

\noindent (iii). A central surface brightness fainter
than $\mu_{B}$ = 24.0 mag arcsec$^{-2}$ and a non-centrally concentrated
light profile that is close to exponential.  These are properties
of nearby dwarf ellipticals and can be used for distinguishing
dEs from giant ellipticals and background systems.  There is some limited 
overlap in luminosity between the giant ellipticals and the LMCGs.

Any objects that do not meet this criteria, and were not previously identified
as stars, are considered background objects.  These morphological 
classifications are the
basis for our decisions concerning cluster membership, and for defining
different galaxy populations.  We did not use the surface-brightness magnitude
relationship to define which objects are in the cluster.  The good fit between
these two parameters is however a check on the reliability of our method for 
picking out real cluster members (see \S 3).
The finally tally is: 160  LMCGs, 28 giant galaxies and 716 background systems.

\section{Background Galaxies}

The misidentification of background galaxies as true cluster members
is a serious problem when
trying to study the faintest members of a cluster.  Popular techniques to
avoid this include subtracting background fields from number
counts, although this method possibly under subtracts  
the number of background galaxies (Valotto, Moore \& Lambas
2001). Another method, used with some success by Secker et al.~(1997),
rejects galaxies redder than the reddest giant elliptical, and
corrects statistically measured properties by using control fields.
Other methods utilize morphological information to determine cluster
membership  (e.g., Binggeli et al.~1985; Ferguson \&
Sandage 1990).   The morphological method is more reliable than one might
think a priori; most ($> 90$\%) of
objects identified as cluster members using this technique  were later
confirmed as members from velocity measurements
(e.g., Binggeli, Popescu \& Tammann 1993; Drinkwater
et al.~2000). By far the most reliable method is to base cluster membership on
measured radial velocities (e.g., Drinkwater et al.~2001).  There are,
however, no clusters where
redshifts are known for all galaxies with 
magnitudes in the range of those in the present study.   The only clusters where radial velocities of
the faintest galaxies (with $M_{B} > -16$) have been obtained are the
nearby Virgo and Fornax clusters (Paper~I; Drinkwater et al.~2001).

\begin{inlinefigure}
\begin{center}
\resizebox{\textwidth}{!}{\includegraphics{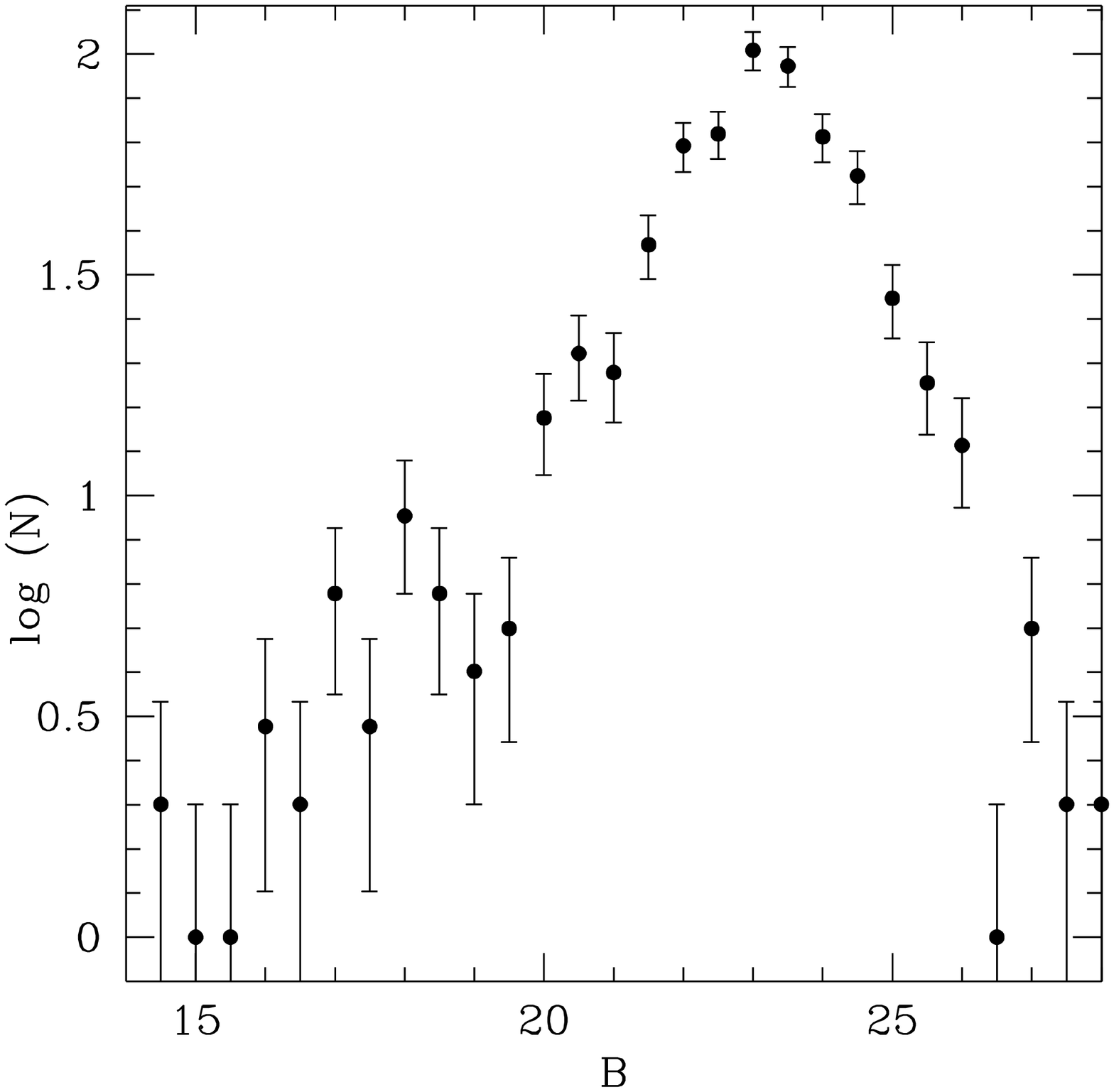}}
\end{center}
\figcaption{Basic luminosity function of all galaxies detected by FOCAS with
sizes R$_{p} > 1$\arcsec\, in our Perseus survey field in the B band.}
\end{inlinefigure}

Background field 
galaxies can often be identified by their sub-structures arising
from spiral arms or other irregularities
produced by star-formation or mergers, which are both
common at high redshift (Conselice
2001).  From a comparison of images of the Hubble Deep Field taken with the 
WIYN telescope with the original HST data, Conselice \&
Gallagher (1999)  determined that images such as the ones used in
the present study (S2kB camera in good seeing) allow  the internal structures 
of moderate redshift  galaxies out to $z \sim 0.5$ to be
detected and resolved.  The problem with distinguishing  background
galaxies from faint cluster members is that higher-$z$ galaxies are
not only small, but they also have a low surface brightness, due to
cosmological dimming, and so can mimic cluster dwarfs.   Since
most high-$z$ galaxies detectable with WIYN are intrinsically of high surface
brightness and have distorted structures, we are able to statistically
separate them from cluster LMCGs; we find only two cases of galaxies
that might marginally be classified as early-type LMCGs in a WIYN image of
the Hubble Deep Field, revealing a very low contamination rate.

To determine if we could
morphologically distinguish galaxy populations at intermediate redshifts, we
simulated R-band images  based on nearby galaxies from Frei
et al.~(1996) redshifted out to  $z \sim 0.5$.  Spiral structures
were easily visible, although the elliptical galaxies proved to be 
much harder to distinguish from
LMCGs.  Ellipticals are however rarely seen in the field or at
high-$z$ outside of clusters, and there is no evidence, prior to this paper,
for any clusters behind the central region of Perseus. We also set a minimum 
size limit in FOCAS for detections of candidate Perseus cluster members, 
of 1\arcsec\, (or $\sim 300$~pc at the distance of the Perseus cluster); 
most galaxies smaller than this are probably in the background. 
We conclude it is likely that actual cluster
members can be distinguished from backgrounds objects through 
identifying substructure, although there will inevitably be some 
misidentifications.

\begin{inlinefigure}
\begin{center}
\resizebox{\textwidth}{!}{\includegraphics{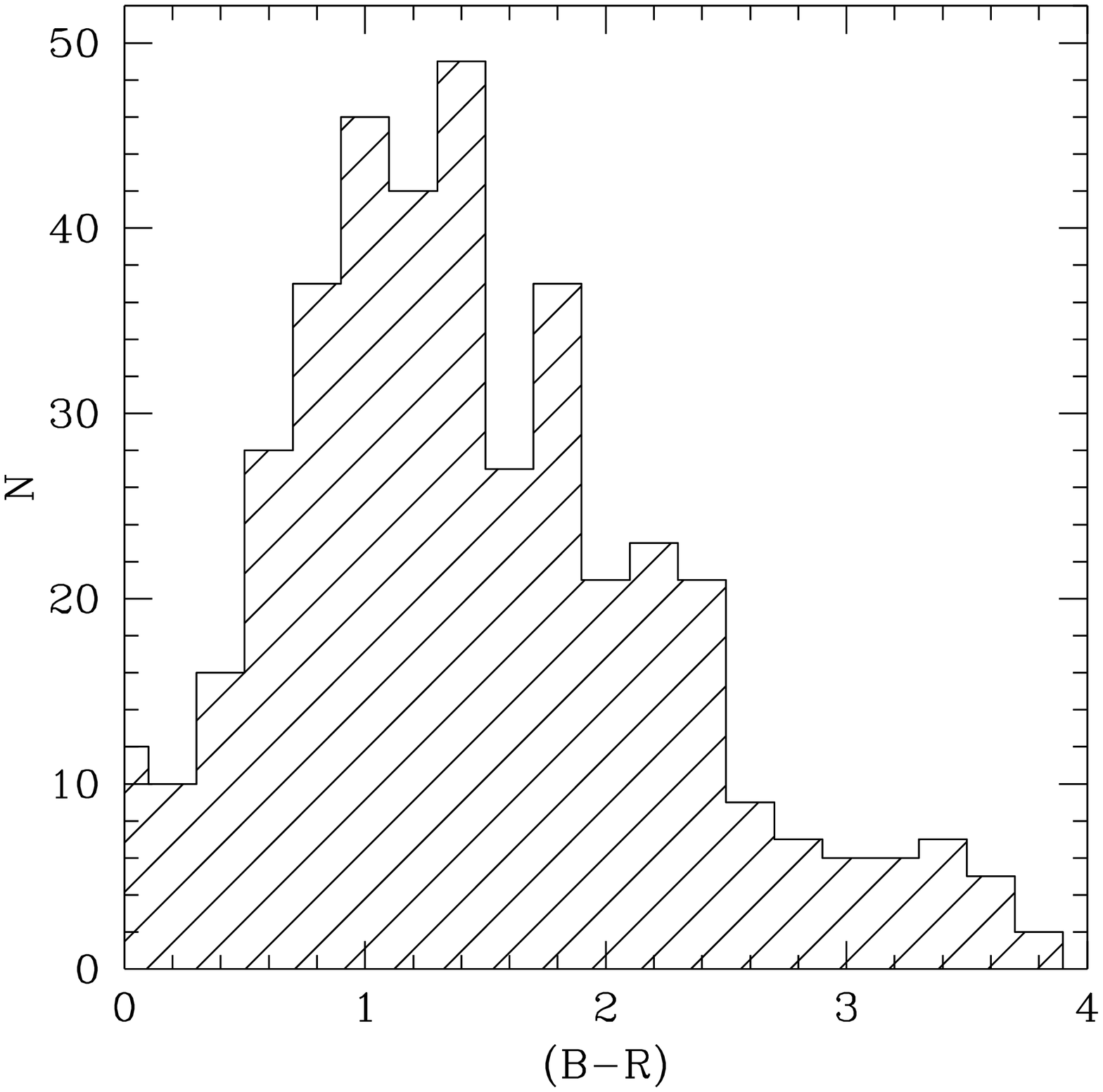}}
\end{center}
\figcaption{Histogram of (B$-$R) colors of morphologically selected
background galaxies seen towards the Perseus cluster center.}
\end{inlinefigure}

Figure~5 shows the resulting color distribution of galaxies in our
images that we have classified as being background objects,
based on their morphologies.  This diagram illustrates a potential
problem with using color selection techniques to remove background
objects, since the typical range of (B--R) color used to assign
cluster membership is $0.9<{\rm (B-R)}<1.8$ (e.g., Secker et al. 1997)
and as seen in Figure~5, many galaxies with colors in this range have
morphologies consistent with their being background objects.  A large
fraction of these galaxies have extremely low S/N ratios in the U
band, which is another indication that they are background objects
since high-$z$ galaxies are usually very faint at shorter observed
wavelengths (e.g., Steidel \& Hamilton 1992).  However, some
misidentifications are unfortunately inevitable.  This color based
rejection technique we and others (e.g., Secker et al. 1997) use will also 
remove populations of faint
cluster galaxies with unusual stellar populations that may exist
(e.g., Drinkwater et al.~2001). From other rich clusters, such as
Virgo, we know that the faint end of the luminosity function is
dominated by structurally smooth early-type LMCGs (Binggeli et al.~
1985). As such, our sample contains a {\em lower} limit to the number
of faint galaxies that might exist in the central region of Perseus as
any irregular appearing objects will be thrown out as being in the
background.

Another indication that
we are rejecting mostly true background galaxies comes from the surface
brightness-magnitude diagram for objects chosen as cluster and background
members (Figure~6).  Figure~6 plots the absolute magnitude for cluster
members, M$_{B}$ vs the central surface brightness, $\mu_{B}$.
While nearly all the cluster members fall along
a well defined correlation of surface brightness with magnitude (\S 4.2),
there is a large spread in position of these background objects on the
same diagram with apparent magnitude plotted vs. surface brightness 
(Figure 6b).  The dashed
curve in Figure~6b shows where elliptical galaxies would be on this
diagram if they were at redshifts higher than that of the Perseus cluster.  The
clump of objects in the upper right of Figure~6b is possibly a background
cluster of galaxies at $z \sim 0.55$ (see \S 4.2 for further discussion). 
Further proof of the success of this method for identifying background
galaxies will require a redshift survey to determine how accurate our 
estimates are.  Another check would be to see if the galaxies
we pick out as background systems are distributed in projected space as would
be expected in the general background field.  If misidentified cluster members
were in this sample then we would see a clustering of galaxies 
towards the center
of the cluster.  While we do not have the area to fully test this, the
background systems do not appear to cluster towards the center of Perseus.

\section{Analysis}

\subsection{The Color-Magnitude Relation}

One of the most remarkable properties of cluster galaxies is the
presence of a well-defined color-magnitude relationship among the
non-dwarf early-type galaxies.  This sequence is now well
characterized and has been observed in several other nearby clusters,
including Coma (e.g., Secker et al.~1997) and Virgo (e.g. Bower, Lucey
\& Ellis 1992).  The color-magnitude relationship in clusters was
first discussed in detail by Sandage (1972) and Visvanathan and
Sandage (1977) for Coma and Virgo.  It is likely a universal
relationship that varies little between different nearby clusters
(Bower et al.~1992).  The color-magnitude relation is also seen in
high redshift clusters, with a shift consistent with passive evolution
between then and now (Stanford, Eisenhardt \& Dickinson~1998).  This is usually
interpreted as a metallicity effect (e.g., Larson 1974), where more
massive galaxies are able to hold on to metals produced in stellar
nucleosynthesis and self-enrich through successive generations of
stars (cf.  Worthey 1994).  The lower mass, and hence likely fainter,
ellipticals have lower escape speeds and suffer enhanced loss of
metal-rich supernova ejecta (e.g., Dekel \& Silk 1986).  This idea is
further suggested by the tight correlation between the strength of the
Mg$_{2}$ absorption line index and the internal velocity dispersions
of early-type galaxies (e.g., Bender, Burstein \& Faber 1993).

We investigated this relationship in the Perseus cluster with our UBR
photometry down to our completeness limit.  We used the $(B-R)_{0}$
color index, which is a powerful physical diagnostic  sensitive
to metallicity in old stellar populations such as globular clusters
(e.g., Harris 1996).  The general color-magnitude trend is confirmed
for galaxies in the Perseus cluster with M$_{B} < -16$ (solid line
in Figure~7; note that the colors are the mean within R$_{p}$).  
A  least squares fit gives  that the relationship between
magnitude and color is: 

\begin{equation}
(B-R)_{0} = (-0.055\pm0.009)M_{B} + (0.456\pm0.16),
\end{equation}

\noindent for objects with M$_{B} < -16$.
The relationship between color and apparent magnitude 
found by Secker et al.~(1997) for galaxies in
the Coma cluster is: $(B-R)_{0} = (-0.056\pm0.002)B_{0} +
(2.41\pm0.04).$

\noindent Converting this into an absolute magnitude B-band relation gives

\begin{equation}
(B-R)_{0} = (-0.056)M_{B} + 0.452.
\end{equation}

\noindent Thus, within the errors, the two clusters have identical
color-magnitude sequences for the brighter galaxies.
The only bright galaxy in the central Perseus cluster region we study
that does not fit along the
color-magnitude relationship is NGC~1275, a bizarre bright and blue galaxy 
with recent
star-formation that is likely undergoing rapid evolution (Conselice et al.
2001b).  It is also the only galaxy over-exposed in both the
mini-mosaic and S2kB images and thus does
not appear on Figure~7.  

As can be seen in Figure~7, there is a large scatter in the
color-magnitude relationship at M$_{B} > -16$.  The rms scatter from the
color-magnitude relation fiducial sequence, calculated only from
galaxies within 1~magnitude of the fiducial to minimize possible
contamination from misidentified non-members, is plotted as a function
of magnitude in Figure~8.  
The scatter is very small at M$_{B} <
-16$, $\sim$ $\sigma$ = 0.07, but rises to $\sigma = 0.54$ at
M$_{B} = -13$.  Out to M$_{B} = -13$, the scatter can be characterized
as

\begin{equation}
\sigma = (0.009\pm0.003) \times 10^{\gamma},
\end{equation}

\noindent where $\gamma$ = (M$_{B}$+23.44$\pm$0.89)/(6.81$\pm$0.68).
\noindent This scatter essentially remains unchanged between M$_{B} = -13$
and M$_{B} = -11$, potentially the result of the color criteria for
membership selection, (B$-$R)$_{0} < 2$, which can remove real cluster members 
that are very red, decreasing the observed scatter at fainter magnitudes.

\begin{inlinefigure}
\begin{center}
\resizebox{\textwidth}{!}{\includegraphics[bb = -5 -5 575 575]{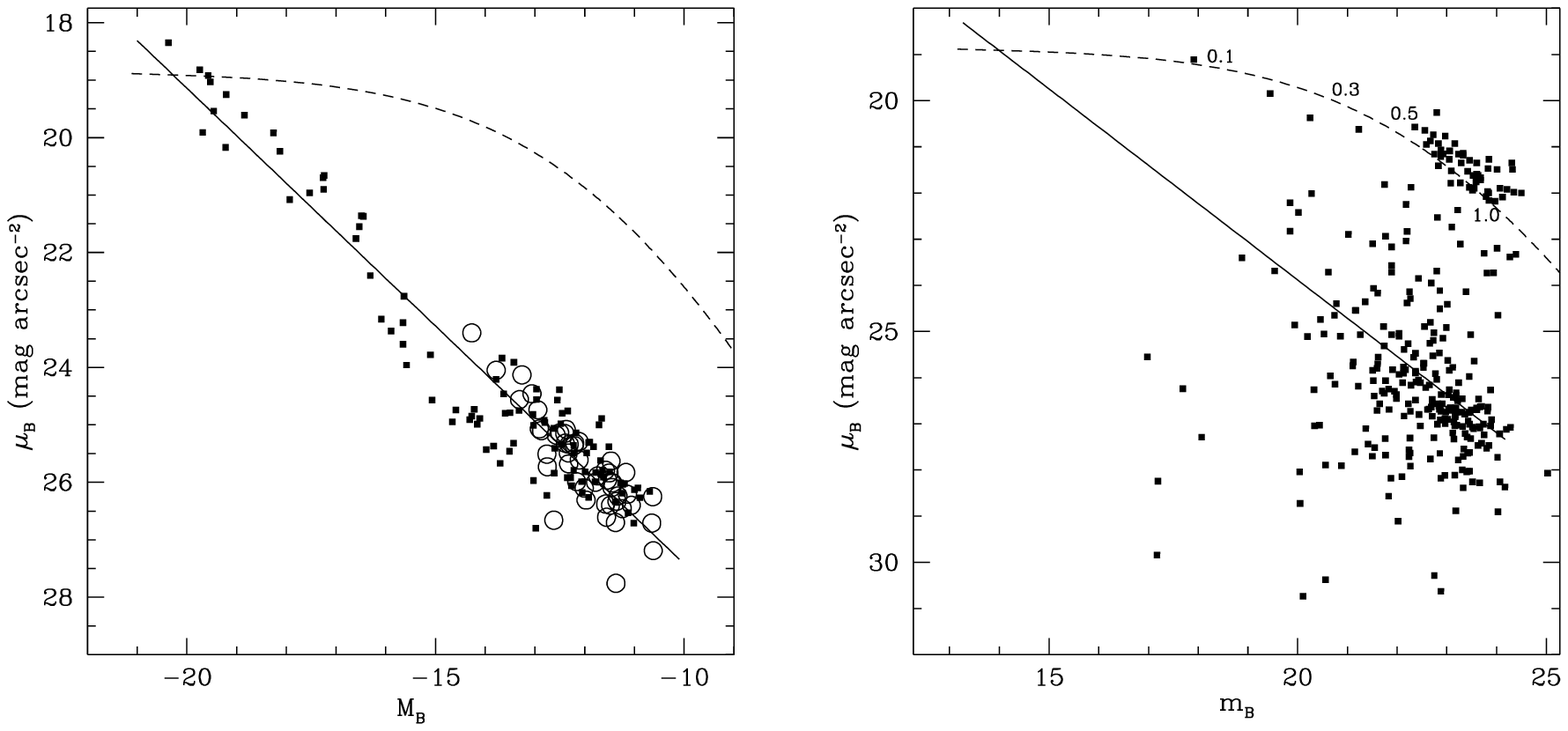}}
\end{center}
\vspace{-4cm}
\figcaption{Absolute magnitude M$_{B}$ versus the central (2\arcsec) surface
brightness $\mu_{B}$ for (a) LMCGs and ellipticals chosen to
be in the cluster. Panel (b) shows the corresponding magnitude-surface
brightness plot for objects 
likely in the background, but plotted against apparent magnitude 
m$_{B}$. The solid small squares in (a) are the 139 objects that are within
2$\sigma$ of the color-magnitude relationship, including the giant 
ellipticals (see text).  The open circles are the 49 objects redder than 
2$\sigma$ of the color-magnitude relationship.
The dashed line shows where a M$_{B}$ = -20
galaxy would be seen on this diagram if it were at redshifts from 0 to 1, 
with z = 0.1, 0.3, 0.5 and 1 labeled on panel (b).
The clustering of points at m$_{B} \sim 23$
at $\mu_{B} \sim 21$ is possibly a background cluster at $z \sim 0.55$.}
\end{inlinefigure}

A similar large amplitude scatter was observed at these faint
magnitudes by Secker et al.~(1997), who used a different approach to
remove background galaxies.  Could this scatter be produced solely from 
foreground or background misidentified as cluster galaxies?
 It is unlikely that many of the blue
objects in this scatter could be stars, or background/foreground
galaxies.  Galaxies at higher redshifts undergoing star-formation are
not as blue as these objects due to large k-corrections.  Since
these objects are clearly resolved they cannot be stars, and they are
unlikely to be foreground field galaxies, since very few field dwarfs
seem to exist.  These objects are also not in any particular part of
our fields, such as near large galaxies, as would be expected for
globular clusters.  Several other groups have recently found a
population of red low-luminosity galaxies in, for example, the Fornax
(Rakos et al.~2001) and Coma clusters (Adami et al.~2000; Mobasher
et al. 2001).  However,
the red objects have a higher chance of being background objects, and
spectroscopy is required to confirm their cluster membership. In Paper
III we discuss in detail the likely physical reasons for this scatter,
and argue more firmly for its existence.

\begin{inlinefigure}
\begin{center}
\resizebox{\textwidth}{!}{\includegraphics{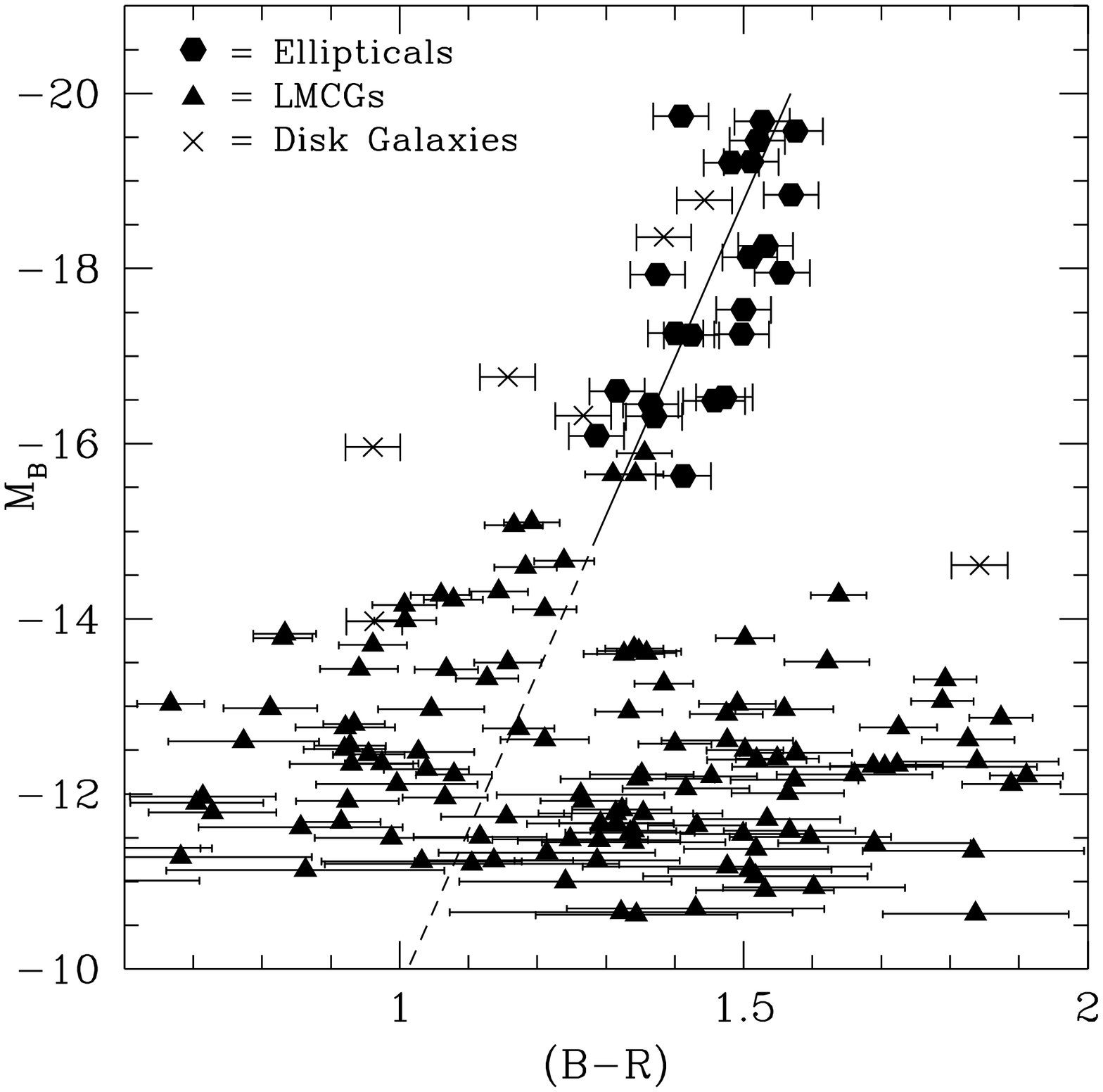}}
\end{center}
\figcaption{Color magnitude diagram of all galaxies identified
as candidate cluster members. The solid line shows the proper fit from
eq. (5) to the giant
ellipticals, and the extension of this correlation to fainter magnitudes
is shown as a dashed line.}
\end{inlinefigure}

A (B--R) vs. (U--B) diagram for those galaxies with UBR photometry is
shown in Figure~9.  Plotted on this diagram are the predictions for
single-age, old populations with metallicities between [Fe/H] = 0.5 to
-2.  The solid line is a 18~Gyr isochrone, the dashed line is a
12~Gyr one, and the dot-dashed line is a 5~Gyr isochrone.  These are 
based on the stellar synthesis models
of Worthey (1994).  In these models we use a standard Salpeter IMF slope of
2.35, with a mass range of 0.1 \solm - 100 \solm,
allowing only passive evolution after an initial single burst of 
star-formation, and adopt a single global metallicity.  This diagram is 
broadly consistent with the interpretations derived from the color-magnitude
relation.  The
ellipticals are located in a very small locus about [(U-B), (B-R)] =
[0.5, 1.6] while the fainter galaxies with M$_{B} > -16$ are generally 
bluer in (B-R) and (U-B) than are the brighter galaxies.  For the most part 
the colors of all galaxies in the central
regions of the Perseus cluster are consistent with containing old ($>$ few 
Gyrs) stellar populations, whose color differences are possibly the result
of variations in metallicities (Paper III).  The lack of a correlation
between color and M$_{B}$ for lower stellar mass cluster members
in combination with colors typical of older, moderate to high
metallicity stellar populations 
suggests that these objects are more metal-rich and hence could have
been, or are, more massive than their luminosity indicates.

\subsection{Surface-Brightness Distributions}

In this section we investigate trends in the surface brightnesses of 
Perseus galaxies.  When comparing to other studies it must however be
remembered that the method we use to define
magnitudes differs from previous studies.  This, however, does not
appear to be important for the photometry, suggesting that the R$_{p}$
radius is a stable choice for measuring proper magnitudes.

\begin{inlinefigure}
\begin{center}
\resizebox{\textwidth}{!}{\includegraphics{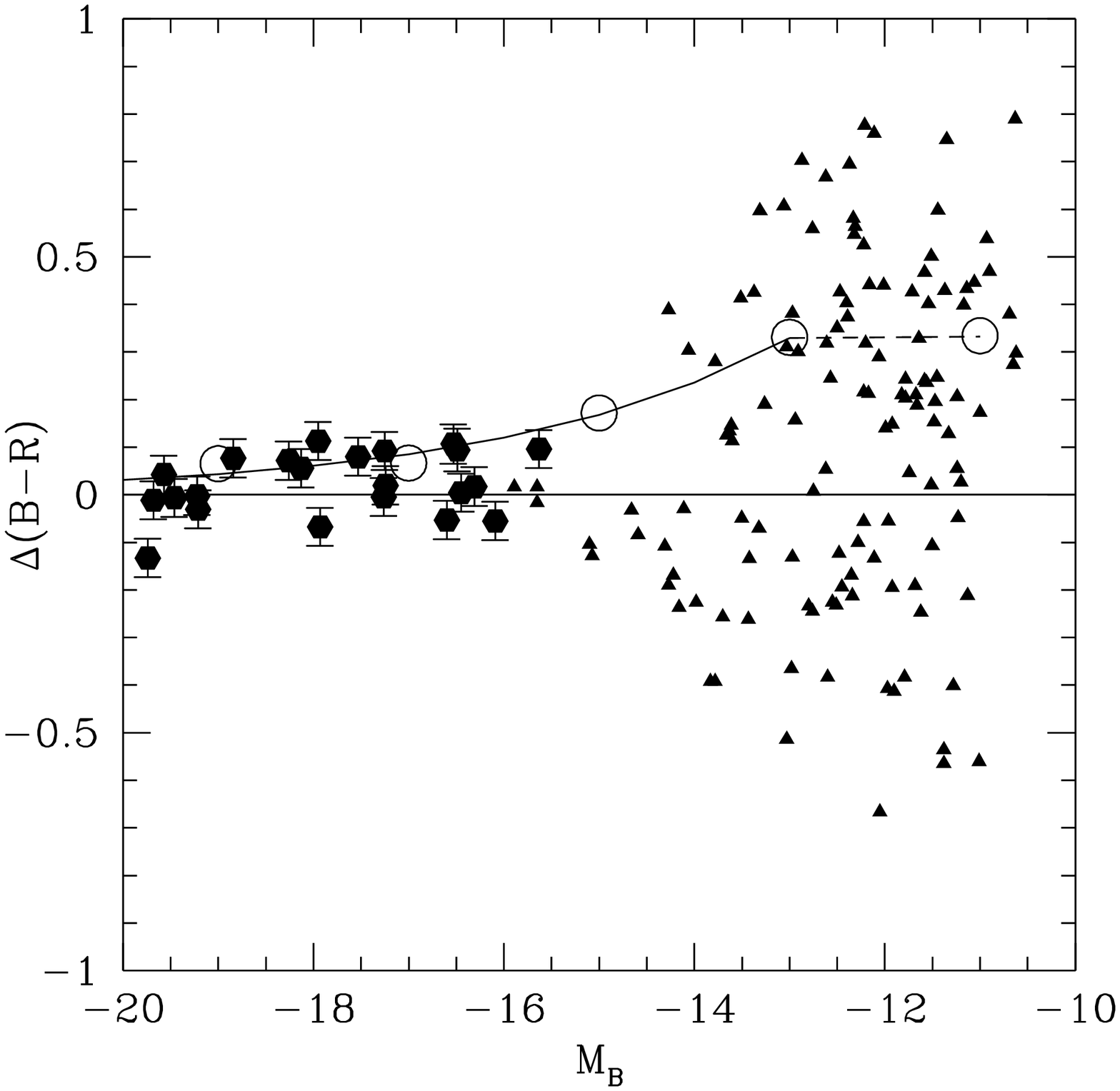}}
\end{center}
\figcaption{Scatter from the color-magnitude relation as a function
of absolute magnitude M$_{B}$.  The open circles are the scatter values
at each magnitude range, while the solid line is a fit to these points
(see text). Symbols are the same as in Figure~7.}
\end{inlinefigure}

The brightest elliptical galaxies in the S2kB images are over exposed.
As such, we use the photometry from the shallower Mini-Mosaic images
to complement the fainter elliptical and LMCG photometry from the S2kB
detector.  Figure~6a
shows the relationship between surface brightness, $\mu_{B}$, as
a function of M$_{B}$ for objects chosen as cluster members.   The 
surface brightness is  measured with a central aperture of 2\arcsec.  This 
figure
demonstrates the well-known fact (e.g., Binggeli \& Cameron 1993) 
that fainter LMCGs have lower surface brightnesses.  The remarkable
feature of this diagram is the relatively tight
correlation between absolute magnitude and central surface brightness over
an 8 magnitude span in brightness. By
fitting this relationship by a least square regression, we obtain

\begin{equation}
\mu_{B} = (0.828\pm0.022)M_{B} + (35.7\pm0.3)
\end{equation}

\noindent with a scatter of $\sigma$ = 0.42 mag arcsec$^{-2}$ at the bright
end, and $\sigma$ = 0.63 mag arcsec$^{-2}$ for the faint galaxies.  Binggeli
\& Cameron
(1991) find that for the Virgo cluster dwarf ellipticals the relationship
$\mu_{B} = 0.75M_{B} + 35.3$ holds, again with  the dispersion
slowly increasing towards the fainter objects. 

However, as opposed to Kormendy (1977) and others, we do not generally
find that the fainter ellipticals have higher surface brightnesses
than the brighter Es.  Either the Kormendy (1977) relationship is not
strong in the central region of the 
Perseus cluster, or as based on our luminosity
function, we do not have enough bright ellipticals in our sample to
see this properly.  In particular, we do not have examples of systems
which are luminous and have low central surface brightnesses (see
e.g., Sandage \& Lubin 2001). The Kormendy relationship has previously
been found to be weakly present in early type cluster members, other
than the most luminous giant ellipticals (Capaccioli \& Caon 1991).
This result is consistent with observations revealing high surface
brightness, `cuspy' centers in moderate luminosity ellipticals and
lower brightness cores in the most luminous objects (Faber et
al. 1997).  We find the same correlation between surface brightness
and magnitude for the early-type LMCGs, as for the giants, when we
plot the mean surface brightness within R$_{p}$, but the LMCGs have a
much larger scatter from the fit between surface brightness and magnitude.

The 49 galaxies which are $>$ 2$\sigma$ redder than the relationship
in eq. (5), which is plotted in Figure~7, are distinguished by open
circles in Figure~6.  Perhaps surprisingly, most of these red objects
fall along the fitted line in Figure~6.  The faint objects that do not
fit as well along this surface-brightness/magnitude relationship are the 
bluer objects (small solid squares), with colors $<2\sigma$ from the 
color-magnitude relationship, which are less likely
than the redder ones to be in the background.  This provides another
indication that the faint red objects are actual cluster members, and
not background galaxies.  If the red objects were ellipticals in, for
example, a background cluster, then the total apparent magnitude
of the galaxy would be very different than if it were near, but the central 
surface brightness should stay relatively constant, short of extreme
cosmological dimming.  These objects would then be above the fitted
relationship (eq.~8).  The dashed curve in Figure 6 shows where on
this diagram a galaxy with M$_{B} = -20$, observed at redshifts in
the range from $z = 0$ to 1, would lie. Clearly, very few objects we
selected as cluster members (Figure~6a) fall anywhere  near this
locus.

Furthermore,  Figure~6b  shows the
surface brightness-magnitude relationship for objects
chosen by our criteria to be in the background.  Besides the obvious fact
that the
scatter about the fit from Figure~6a (the solid line) is large, there
are objects with  surface brightnesses several magnitudes higher
than what would be predicted for LMCGs, if they are dwarf-like,
based on their magnitudes. 
Near m$_{B} = 23$ and  $\mu_{B}$ 
= 21, lying on the dashed curve, there appears to be a cluster of galaxies.   
If this is a background cluster, its location on this diagram 
is consistent with $z \sim 0.55$.   The brightest galaxies
that compose this clump have measured (B$-$R) colors consistent with this
redshift,  assuming they are ellipticals (e.g., Fukugita, Shimasaku,
\& Ichikawa 1995).  Objects that compose this clump cluster on the sky, 
with most of the them in a region of area $\sim$ 3.5 arcmin$^{2}$ centered
roughly at (J2000) 03:18:49, +41:33:30.   For H$_{0}$ = 
70~km~s$^{-1}$~Mpc$^{-1}$, $\Omega_{\rm M}$ = 1.0 gives a diameter of 0.59~kpc
for this candidate cluster, or 0.72~kpc for H$_{0}$ = 
70~km~s$^{-1}$~Mpc$^{-1}$, 
$\Omega_{\rm M}$ = 0.3 and $\Omega_{\lambda}$ = 0.7 both of which are 
reasonable cluster sizes (e.g., Sheldon et al. 2001).

\subsection{Structural Parameter Results}

Structural parameters provide quantitative information concerning
morphologies and properties of galaxies otherwise unobtainable
from pure photometry or spectroscopy. In this section we examine the 
concentration and asymmetry indexes calculated for
cluster members, as well as various measurements of radius.  
Measuring structural parameters
of galaxies requires much better resolution and higher
signal to noise ratios than those needed to acquire accurate photometry.
Because of this, and
to limit the effects of any potential contamination from background
objects, we make further restrictions in the definition of the sample used to
investigate structural parameters.
We only examined galaxies
with effective (within the half-light radius) surface brightness
$\mu_{B} < 26$ mag arcsec$^{-2}$,
and colors of 0.6 $<$ (B$-$R) $<$ 1.65.  This minimizes contamination from
foreground and background objects that are
likely to be, in some form, present in the total sample.

\begin{inlinefigure}
\begin{center}
\resizebox{\textwidth}{!}{\includegraphics{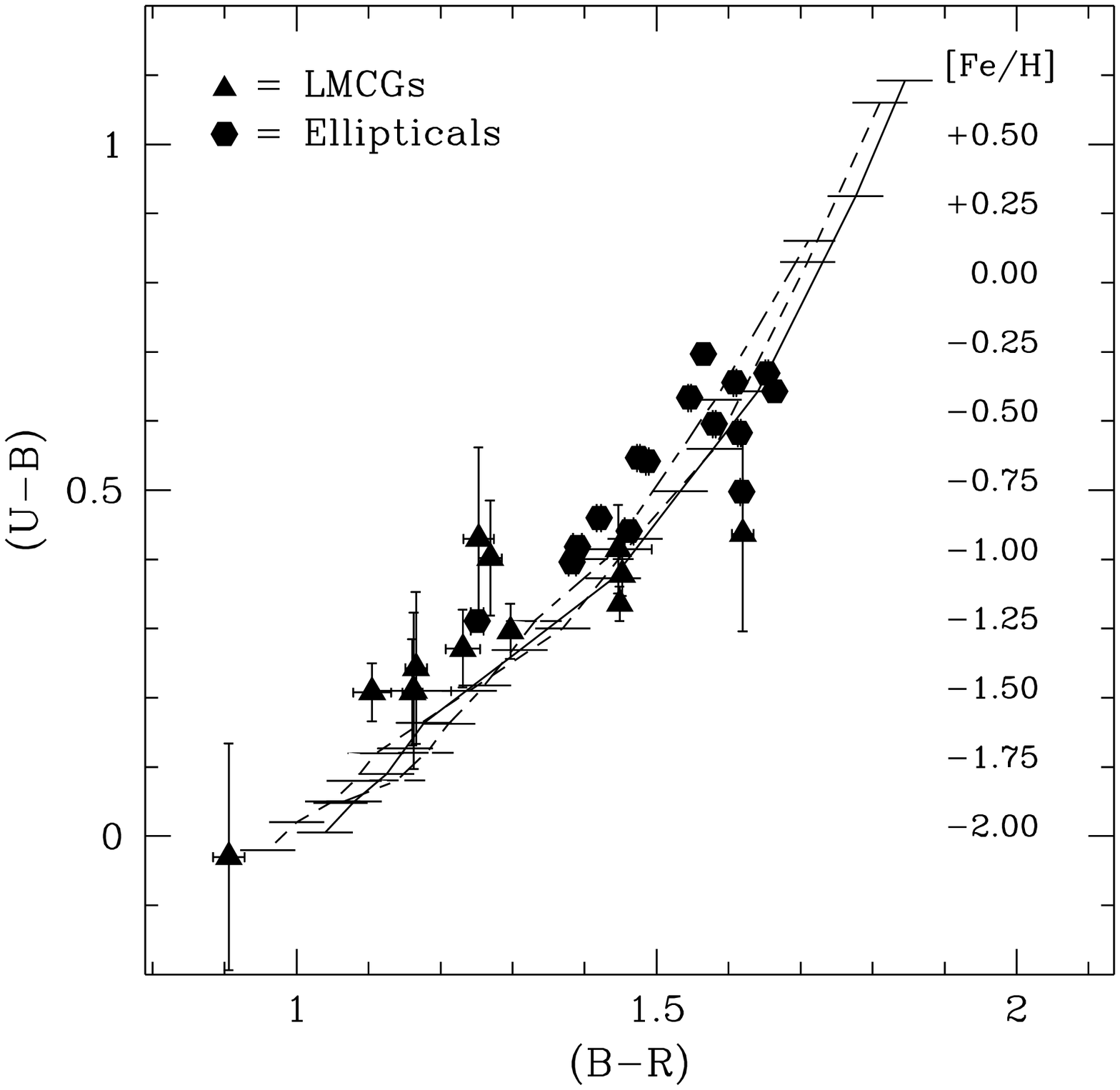}}
\end{center}
\figcaption{UBR color-color diagram of galaxies with U band
photometry.  The dot-dashed, dashed and solid lines are color isochrones of 
5, 12 and 18~Gyr instantaneous burst stellar populations with various 
metallicities
labeled towards the right and denoted as dashed lines on the various
isochrones.  The brightest elliptical galaxies have metallicities slightly 
higher than solar.}
\end{inlinefigure}

\subsubsection{Asymmetry}

The asymmetry structural parameter  (see also \S 2.5) is an effective 
measurement of the merging, interacting, and star-formation properties of
galaxies (Conselice et al.~2000a,b).   Briefly, galaxies
with low asymmetries and red colors are usually classified as ellipticals,
while those with higher asymmetries and bluer colors are disk and irregular
galaxies.  Galaxies with the highest asymmetries are consistent
with interactions or mergers (see Conselice et al.~2000a for a full
discussion).  After computing the asymmetries of all galaxies, we find the
not too surprising result that there is a lack of systems
with high asymmetries.  In its inner $\sim$173 arcmin$^{2}$ ($\sim$0.1
Mpc$^{3}$) the Perseus cluster has only one galaxy consistent with an
ongoing major merger, NGC~1275 (Conselice et al.~2001b).

For the most part, cluster giants
are found to be consistent with classical ellipticals,
with low asymmetries and red colors, as might have been expected.  
However, we see something
different for the LMCGs.  Among field galaxies there is a
significant trend between asymmetry and $(B-V)_{0}$ color,
a good indicator of recent star-formation (Conselice et al.~2000a).
In the Perseus cluster we find no correlation;
bluer galaxies, which are LMCGs, are generally symmetric.

Two possible effects are responsible for this lack of an observed
correlation between color and asymmetry.  The first is that the
fainter LMCGs, which are small and blue objects, might not be
sufficiently resolved to show any clumpy regions of star-formation
(e.g., Gallagher \& Hunter 1989).  Conselice et al.~(2000a) performed
simulations on nearby galaxies and determined that a reliable
asymmetry can be obtained in galaxies with structures resolved on
scales greater than 500~pc.  In Perseus, we are resolving scales down
to 210~pc.  These simulations were however done on large spiral and
elliptical galaxies, and they might not be applicable to small
early-type LMCGs which may have structures smaller than those found in larger
galaxies.

The other possible explanation is that (B$-$R) is only measuring 
metallicity, and little to no recent star-formation is present in these
systems. This would imply that all
of these faint galaxies have been quiescent for several billion
years.  Use of a color index that is  less sensitive to metallicity would 
then produce a narrower spread in color, and maintain the asymmetry spread. 

We conclude from this that most galaxies in the central regions of
the Perseus
cluster contain old stellar populations, with little to no recent 
star-formation and (B$-$R) colors driven primary by metallicity.  This is  
consistent with a lack of H$\alpha$ emission
in any galaxy in the central regions of the Perseus cluster, besides the
giant elliptical NGC~1275 (Conselice et al.~2001a), and with the UBR
diagram (Figure~9) showing the colors of LMCGs match those of old
stellar populations with differing metallicities. In Paper~III we examine the
color maps of the Perseus LMCGs to determine if any stellar population
gradients are present.

\subsubsection{Radii and Light Concentrations}

How do the sizes of Perseus cluster galaxies correlate with other
features?  Figure~10 shows the relationship between the half-light (or 
effective) radii, R$_{e}$, defined in \S2.5, and the total B-band magnitude. In
general, brighter galaxies have larger half-light radii; a 
least-squares fit to all galaxies gives the following correlation: 
$$ R_{e} ({\rm kpc}) = (-0.1 \pm 0.008) \times M_{B} - (1.0\pm0.1),$$

\noindent which is plotted as the solid line in Figure~10. A large amount of 
the scatter about this relation arises from the disk galaxies. If we consider 
just the LMCGs then the slope of this correlation becomes
steeper.  The formal fit restricted to LMCGs is given by: 
$$R_{\rm e, LMCG} ({\rm kpc}) = (-0.2 \pm
0.02) \times M_{B} - (2.3\pm0.2),$$ and is shown as a dashed line in Figure~10.
These are similar
to the relationships found for dwarfs and giants in the Virgo cluster
by Binggeli \& Cameron (1991).

The concentration
index (as defined 
in \S 2.5) is measured from the curve of growth of each galaxy. 
The 
giant ellipticals are the most concentrated objects, with C $\sim$ 3.
The LMCGs have a wide range of concentration values, but in general they are
the least concentrated objects, demonstrating perhaps a fundamental difference
in formation mechanisms.

\begin{inlinefigure}
\begin{center}
\resizebox{\textwidth}{!}{\includegraphics{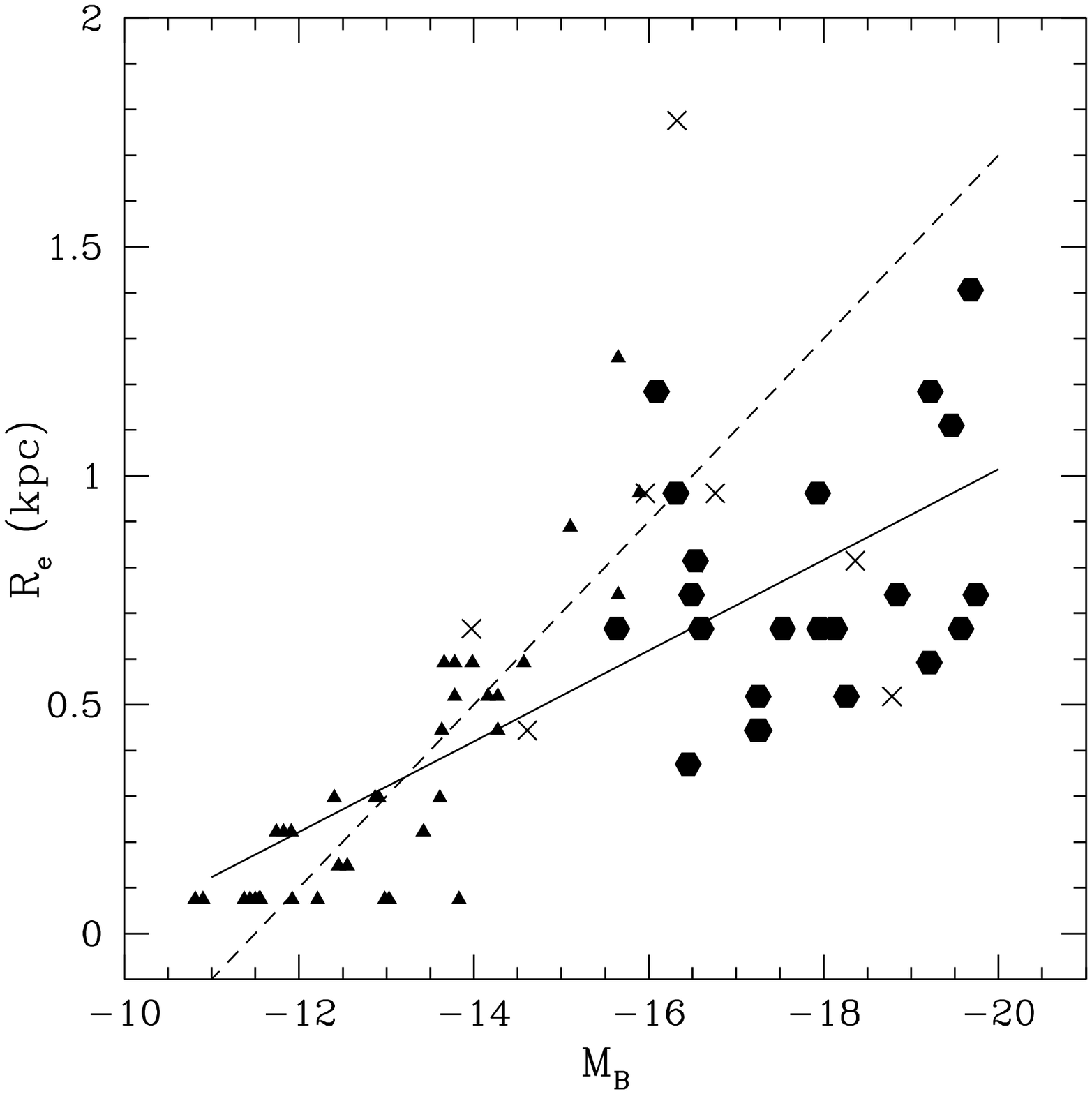}}
\end{center}
\figcaption{Effective radii R$_{\rm e}$ as a function of
absolute magnitude M$_{B}$.  Symbols are the same as in Figure 7.  The
solid line is a proper fit to all of the data (see \S 4.3.2), while the
dashed line is a fit to the LMCGs alone.}
\end{inlinefigure}

The low concentration values of the LMCGs are another piece of evidence that
these objects are cluster members, and not background galaxies.  Galaxies at
redshifts $z>0.1$
typically have intrinsically steep radial profiles if they
are ellipticals, or bulge dominated systems. Early-type background galaxies 
have much higher concentration indexes than those found for a majority of
the LMCGs.  If background galaxies have pure exponential profiles then
they could in principle be misidentified as LMCGs, however these objects
are usually too faint and small to be included in the our sample. 

\subsection{The Perseus Cluster Central Luminosity Function}

 The number of galaxies per unit magnitude interval is a useful observational
constraint to compare galaxy formation scenarios with observational data (e.g.,
Press \& Schechter 1974).
A popular method for doing this is to compute luminosity functions (LF)
in different environments, especially in clusters.  One major
reason for fitting luminosity functions is to determine the relative
number of low-mass systems in comparison to larger galaxies by
measuring the value of the faint-end slope, $\alpha$ which has measured
values between $\sim$ -1.0 and -2.3 (e.g., Thompson \& Gregory 1993; Biviano
et al. 1995; Bernstein et al. 1995; Lopez-Cruz et al. 1997; Secker
et al. 1997; Phillips et al.
1998; Trentham 1998; De Propris \& Pritchet
1998; Adami et al. 2000; Beijersbergen et al. 2002).  The typical way to 
fit luminosity functions is to  model the data with some
parameterized fit, such as a power-law or the Schechter (1976)
function which has the form:

\begin{equation}
\phi(L) dL = \phi^{*}(L/L^{*})^{\alpha} \times {\rm exp}(-L/L^{*}) \frac{dL}{L^{*}},
\end{equation}

\noindent where $\phi(L) dL$ is the number of galaxies at luminosity $L$ within
the interval $dL$. The parameters $\phi^{*}, L^{*}$ and $\alpha$ are
the normalization, characteristic luminosity and faint-end slope for
the luminosity distributions.  The Schechter function can be written
in terms of magnitudes as,

\begin{equation}
\phi(M) = (0.4 \times {\rm ln} 10) \phi^{*} [10^{0.4(M^{*}-M)}]^{1+\alpha} {\rm exp}[-10^{0.4(M^{*}-M)}].
\end{equation}

\noindent Previous to this study, De Propris \& Pritchet (1998) obtained a
luminosity function for the central regions of the Perseus cluster, finding a faint-end slope $\alpha =
-1.56\pm0.07$ within the magnitude range $-19.4 < M_{I} < -13.4$.
The Coma cluster is generally observed to have a similar
faint end slope of $\alpha = -1.4$ (e.g., Thompson \& Gregory 1993;
Biviano et al. 1995; Bernstein et al.~1995;
Secker et al.~1997; Lopez-Cruz et al. 1997; Beijersbergen et al. 2002).  

The WIYN images used here are deeper, by 1 magnitude, than the data
used in De Propris \& Pritchet (1998), assuming a typical color of
B$-$I = 1.6 for the lowest luminosity galaxies.  As was shown in \S 3,
we are likely incomplete at magnitudes fainter than M$_{B} =
-10.7$.  Figure~11 shows the resulting luminosity function to this
limit, after removing all likely background candidates.  We fitted the
Schechter function and a power law, of the form dN/dL = L$^{\alpha}$, to
these number counts by using a weighted $\chi^{2}$ maximum-likelihood
minimization method.  The resulting fit to the number counts is shown in
Figure~11 as a solid line.  The faint-end slope is computed as
$\alpha$ = $-1.44\pm$0.04, with M$^{*}$ fixed to $-19$, although
note that the value of the slope $\alpha$ is not highly
dependent on the choice of M$^{*}$.  Fitting the luminosity function to a 
power-law gives a slope of $\alpha$ = $-1.42\pm0.03$.  These values are
slightly flatter than the slope found by De Propris \& Pritchet (1998).

This suggests that perhaps in addition to a universal color-magnitude
relation for bright cluster galaxies, there is a consistent process at
work to produce the same faint luminosity functions in some rich clusters.
Differences between galaxies in clusters seem to be limited to an increased 
dispersion in the color-magnitude relation for the faint galaxies, and an 
observed variations in the structural and
evolutionary properties of the brightest cluster galaxies.  It appears that 
the faintest and very
brightest galaxies in clusters are undergoing the most active evolution over 
the last few Gyrs (Conselice et al.~2001a,b).  Paper III will address, among
other things, possible scenarios for the formation of these 
low-mass galaxy systems.

\subsection{Properties of Unusual Perseus Galaxies}

Our high quality WIYN images allow an unprecedented opportunity to
study individual galaxies in the Perseus cluster 
(cf.~Conselice et al.~2001b for NGC~1275). 
We examine the detailed
properties of the low-mass Perseus galaxies in Paper III;  there are,
however, two interesting examples of unusual galaxies that we briefly
discuss here.
\begin{inlinefigure}
\begin{center}
\resizebox{\textwidth}{!}{\includegraphics{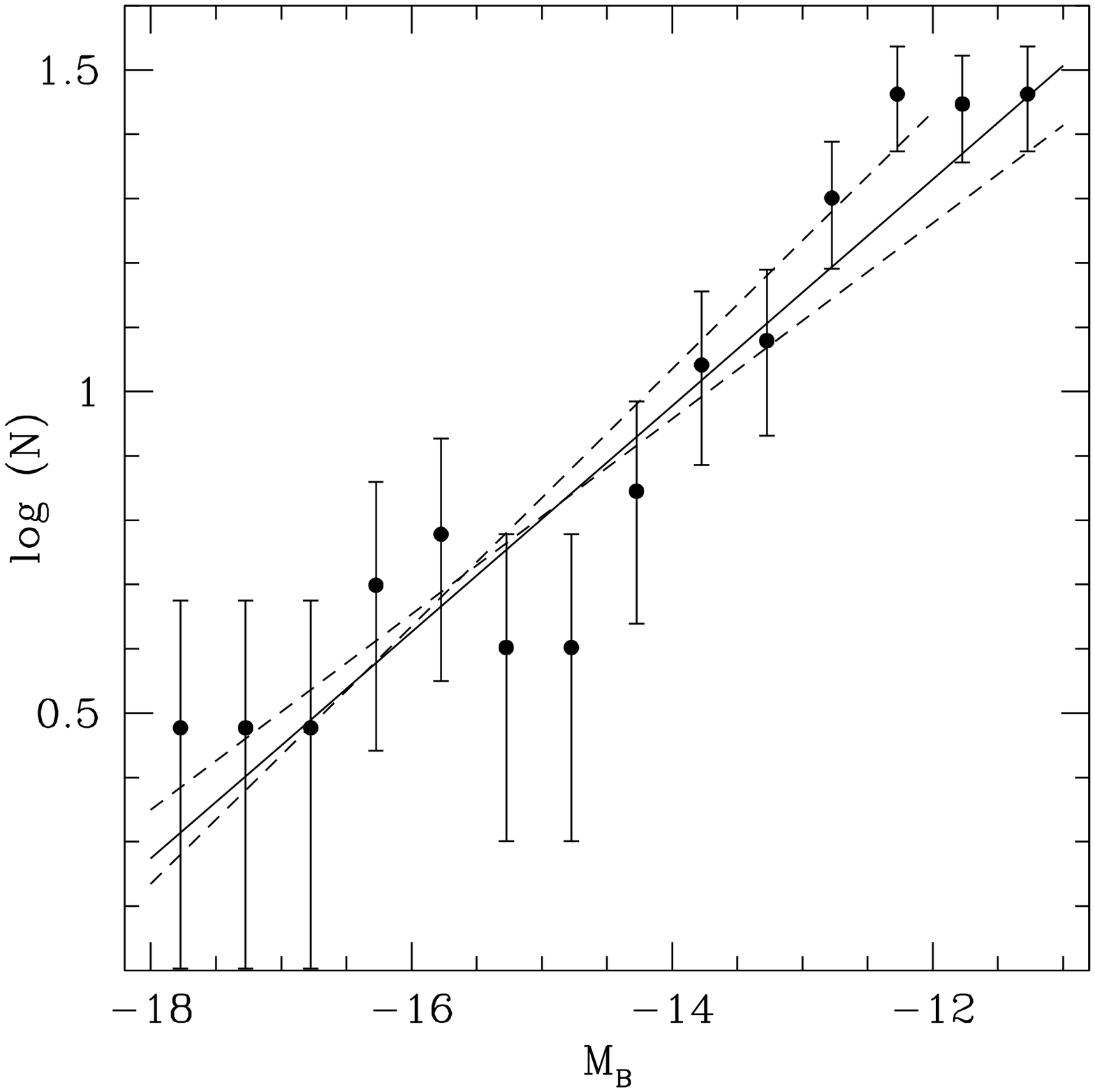}}
\end{center}
\figcaption{The luminosity function for galaxies with M$_{B} > -18$ in the central
region of the Perseus cluster.
The solid line is the proper fit, while the dashed lines are 1$\sigma$
variations on this fit.}
\end{inlinefigure}

After examining the outer isophotes of all ellipticals in the sample,
we noticed that one, located at a projected distance of 110~kpc from
NGC~1275, shows isophote twists (Figure~12b).    Elliptical isophotes are fit 
to this galaxy by
using the Fourier fitting series described in Jedrzejewski (1987).
For comparison, Figure~12a shows a normal, typical Perseus elliptical galaxy, 
its ellipticity profiles (defined as 1$-$b/a), and the positional
angle of its major axis as a function of distance from its center.
The position angle remains roughly constant with radius until
$\sim$~20\arcsec, when the ellipticity becomes too low to produce an
accurate measurement.  The change in position angle of the isophotes
for the galaxy in Figure~12b indicate either a triaxial shape, or that
the galaxy has intrinsic twists that are perhaps remnants of
interactions or from disk galaxy mergers that formed the elliptical
(Gerhard 1983).  This is the only elliptical with this property in the
central region of the Perseus cluster. The lack of twists in 
general is another indication
that giant cluster galaxies are a long-established 
population, since otherwise one might expect  structural
remnants would remain from recent mergers (e.g., Gerhard 1983).  The
far right column in Figure~12 shows the coefficients a$_{4}$ of the 
Fourier fits to the isophotal shapes.  
This component reveals if the isophotes deviate from
pure ellipses.  All three galaxies shown have an average a$_{4}$
$\sim$ 0, indicating that the structures are not dominated by boxy or
disky isophotes.

The galaxy at the bottom of Figure~12 is an unusual object only
$\sim$35~kpc projected distance from NGC~1275. 
This galaxy is located
at (J2000) 03:19:39.6, +41:31:04 and is called SA0426-002 in Conselice
\& Gallagher (1999).  As can be seen, this
galaxy has a butterfly shape, with low surface brightness
`wings', with $\mu_{B}$~=~27~mag~arcsec$^{-2}$.  The core of the
galaxy is also unusual; it is a linear, almost bar-like object, that
gets broader with radius.  This strange object is perhaps the result
of a dynamical interaction with NGC~1275, with the Perseus cluster's
potential or with some other individual galaxy.  It
could also be an object seen in an unusual projection, such as a disk galaxy
with an outer ring that has had its spiral arms removed by tidal
effects.  Figure~12c shows the surface brightness profile of
this galaxy, together with the best-fit  Sersic profile of the form 
$$ I= I_{0}~\times~{\rm exp} [-(r/r_{0})]^{(1/{\rm n})}.$$

\noindent This best-fit profile, with n = 1.19$\pm$0.08, is close to a pure
exponential. This galaxy has a rather blue color, (B$-$R) = 1.3, and faint 
absolute magnitude, M$_{B} = -16.3$.  This is faint enough for
this object to be a type of dwarf galaxy.  
This galaxy  deserves further
attention, since no galaxy known to the authors has a similar
morphological appearance.
\section{Summary}

In this paper we present quantitative techniques for analyzing
deep images of galaxy clusters to investigate galaxy evolution.  We 
illustrate this using new photometric data, obtained with the
WIYN 3.5-m telescope, for  galaxies with M$_{B} < -11$ in 
the central regions of the nearby rich galaxy cluster
Abell 416 (Perseus). We demonstrate that with the use of non-biased detection,
photometric, and structural analysis techniques, we can reduce
background galaxy contamination.  Using this selected sample of
cluster members, we decipher the demographics of central members of the
Perseus cluster.  The main observational results are:

\noindent (i) There are two separate early-type galaxy
`populations' in the central region of Perseus based on photometric scaling 
relationships -- 
a bright one at $M_{B} < -16$, and a distinct fainter one at $M_{B} > -16$
containing early-type low-mass cluster galaxies (LMCGs, defined in \S 1).  
There is a considerable
scatter among LMCGs from the `universal' color-magnitude relation,
extrapolated from a fit to the bright ellipticals, for galaxies with
$M_{B} > -16$. The scatter continues to increase at fainter magnitudes up to
$M_{B} = -13$, a point we discuss further in Paper~III of this series.

\noindent (ii) The number counts of galaxies in the Perseus central
region can be fit by
a Schechter function with a faint-end slope  $\alpha =
-1.44\pm0.04$, close to values found in other clusters.
The color-magnitude relation is found to have a slope, for
galaxies brighter than M$_{B} = -16$, of (B$-$R)/M$_{B}$ = 0.055$\pm$0.009,
similar to that found for Coma (Secker et al.~1997).  We find that
surface brightness scales with luminosity for all early-type cluster
members.  Taken as a group, the photometric properties of Perseus
cluster early-type LMCGs are indistinguishable from those of dwarfs
in the less rich Virgo and Fornax clusters, or in the richer Coma cluster.

\noindent (iii) Aside from the complex NGC~1275 system (Conselice et al.
2001b), we find few unusual galaxies, or galaxies undergoing
rapid evolution in the Perseus cluster core.  Only two galaxies have
evidence for a non-passive evolution during the last few Gyrs.   The Perseus
cluster center therefore contains galaxies that are mostly relaxed
and composed of old stellar populations that
were in place several Gyrs ago.
These results imply that rich cluster galaxy formation is largely consistent
with an early formation and later passive evolution
for {\em the most massive galaxies} (Springel et al. 2001).
There appears to be a possible separate process for the faint cluster
galaxies, as they do not follow the same color-magnitude scaling 
relationship.
Despite this, the similar faint-end luminosity slopes and mean LMCG
color-magnitude relations in nearby clusters suggests that whatever
process is responsible for creating the faintest galaxies in the center
of Perseus is possibly occurring in other clusters.  On the other hand
it must be kept in mind that Perseus is not a typical rich cluster given
its very high galaxy density and its other extreme properties that might make 
galaxies in its center region unique from other clusters.  Similar analyses of
galaxy populations covering a wide range in 
luminosity in other clusters are required to determine this.

We thank the staff of WIYN for their support of these observations
and the referee for helpful comments.  This research was supported
in part by the National Science Foundation (NSF) through grants AST-9803018
to the University of Wisconsin-Madison and AST-9804706 to Johns Hopkins
University. CJC acknowledges support from a Grant-In-Aid of Research
from Sigma Xi and the National Academy of Sciences (NAS) as well as a
Graduate Student Researchers Program (GSRP) Fellowship from NASA and from
the Graduate Student program at the Space Telescope Science Institute (STScI).
RFGW acknowledges a Visiting Fellowship from the UK PPARC.

\begin{inlinefigure}
\begin{center}
\vspace{1cm}
\resizebox{\textwidth}{!}{\includegraphics{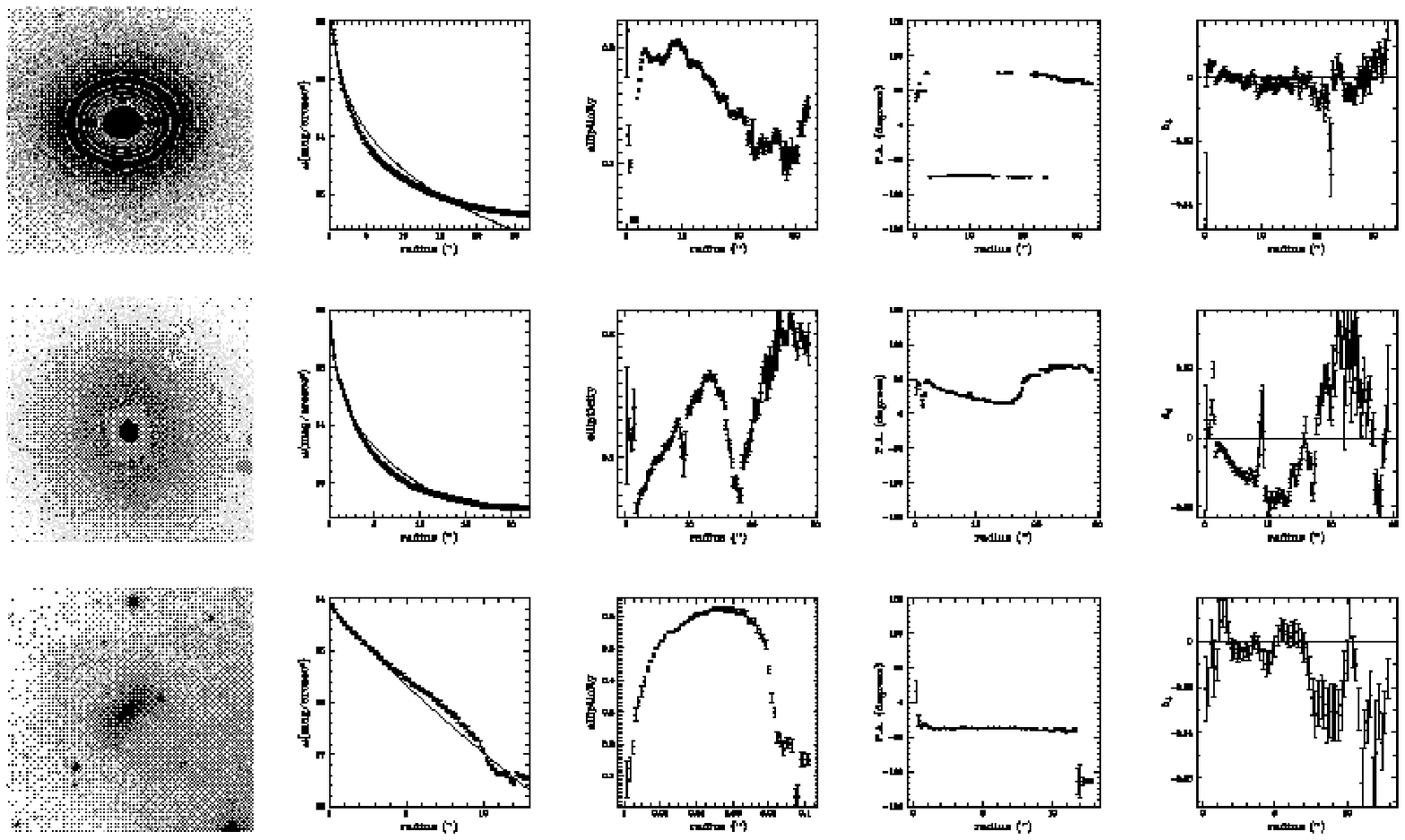}}
\end{center}
\vspace{1cm}
\figcaption{Sample of galaxies in the center of Perseus:
(a) a normal Perseus elliptical galaxy, (b) an elliptical with twisting
isophotes, (c) a  symmetric
galaxy with a peculiar morphology.  The first column shows the image of
each galaxy, while the remaining columns are the surface brightness profiles
with the best Sersic fit shown, and ellipticity, position angle, and a$_{4}$ 
component as a function of radius.}
\end{inlinefigure}

\end{document}